\documentstyle[11pt,epsf]{article}
\newcommand{\beq}{\begin{equation}}
\newcommand{\eeq}{\end{equation}}
\newcommand{\beqa}{\begin{eqnarray}}
\newcommand{\eeqa}{\end{eqnarray}}
\newcommand{\beqan}{\begin{eqnarray*}}
\newcommand{\eeqan}{\end{eqnarray*}}
\newcommand{\no}{\nonumber}

\newcommand{\ul}{\underline}
\newcommand{\ol}{\overline}
\newcommand{\ra}{\rightarrow}

\newcommand{\ben}{\begin{enumerate}}
\newcommand{\een}{\end{enumerate}}
\newcommand{\bfl}{\begin{flushleft}}
\newcommand{\efl}{\end{flushleft}}
\newcommand{\ba}{\begin{array}}
\newcommand{\ea}{\end{array}}
\newcommand{\btab}{\begin{tabular}}
\newcommand{\etab}{\end{tabular}}
\newcommand{\bit}{\begin{itemize}}
\newcommand{\eit}{\end{itemize}}

\newcommand{\cA}{{\cal A}}
\newcommand{\cB}{{\cal B}}

\newcommand{\cN}{{\cal N}}

\newcommand{\vs}{\vspace}
\newcommand{\hs}{\hspace}
\newcommand{\ld}{\lambda}
\newcommand{\G}{\Gamma}

\def \sam {\triangle}
\newcommand{\prepr}[1] {\begin{flushright} {\bf #1} \end{flushright} \vskip
1.cm}
\newcommand{\titul}[1] {\begin{center}{\Large {\bf #1 } } \end{center}
\vskip 0.8cm}

\newcommand{\autor}[1] {\begin{center} \large {\bf \lineskip .3cm #1  }
                        \end{center} }

\newcommand{\lugar}[1] {\begin{center}  {\normalsize \bf \it #1   }
\end{center}}
\newcommand{\abstr}[1] {{\begin{center} \vskip .5cm {\bf \large Abstract
                        \vspace{0pt}} \end{center}}\begin{quote} \small #1
                        \end{quote}}
\newcounter{muni}

\input epsf
\begin{document}
\vspace{.1cm}
\hbadness=10000
\pagenumbering{arabic}
\begin{titlepage}
{\bf \em 29 May, 1995} \hs{80mm}
\prepr{(Revised Version) \\
Preprint PAR/LPTHE/95-09 \\
hep-ph/9503326
 }
\titul{\large A MODEL FOR THE CHARMED $D_s^{+}$ MESON \\
DECAYS INTO THREE PIONS }
\autor{ M. Gourdin\footnote{\rm Postal address: LPTHE, Tour 16, $1^{er}$ Etage,
4 Place Jussieu, F-75252 Paris CEDEX 05, France. \\ . \hs{5mm}
Electronic address : gourdin@lpthe.jussieu.fr},
Y. Y. Keum\footnote{\rm Postal address: LPTHE, Tour 24, $5^{\grave{e}me}$
Etage,
 2 Place Jussieu, F-75251 Paris CEDEX 05, France. \\ . \hs{5mm}
 Electronic address : keum@lpthe.jussieu.fr}
and X. Y. Pham\footnote{\rm Postal address: LPTHE, Tour 16, $1^{er}$ Etage,
4 Place Jussieu, F-75252 Paris CEDEX 05, France. \\ . \hs{5mm}
Electronic address : pham@lpthe.jussieu.fr} }

\lugar{Universit\'e Pierre {\it \&} Marie Curie, Paris VI \\
Universit\'e Denis Diderot, Paris VII \\
Physique Th\'eorique et Hautes Energies \\
Unit\'e associ\'ee au CNRS : D 0280
}

\vs{-14cm}
\thispagestyle{empty}
\vs{133mm}
\newpage
\noindent
\abstr{
\hs{10mm}
We propose a phenomenological three component model
describing the decay amplitude of the process $D_{s}^{+} \ra 3\pi$.
The first component is a constant background,
the second one is a Breit-Wigner amplitude
associated to a quasi two-body $f_{0}(980) \pi^{+}$ state
and the third one another Breit-Wigner amplitude
corresponding to a possible quasi two-body $\rho(770) \pi$ state.
We show that it is possible to reproduce
the observed total rate for
$D_{s}^{+} \ra \pi^{+}\pi^{+}\pi^{-}$
as well as the two other measured branching ratios
for the non resonant part and the resonant $f_{0}\pi^{+}$ one.
Implication of the $\rho\pi^{+}$ state, of which an experimental limit
has been given, is discussed.
An upper bound in the 10 MeV range for the decay constant
$f_{\pi'}$ of the $\pi(1300)$ meson is obtained .

\hs{10mm}
Predictions are given for the
$D_{s}^{+} \ra \pi^{0}\pi^{0}\pi^{+}$ rate
as well as for the  $\pi^{+}$
and $\pi^{-}$ ($\pi^{0}$ and $\pi^{+}$)
energy distributions for these two decay modes
$D_{s}^{+} \ra \pi^{+}\pi^{+}\pi^{-}$
 ($D_{s}^{+} \ra \pi^{0}\pi^{0}\pi^{+}$) respectively.

\large
\vs{10mm}

}
%

\end{titlepage}


\newpage
\hspace{1cm} \large{} {\bf I \hs{3mm} Introduction}
\vspace{0.5cm}
\normalsize

We are interested, in this paper, in the decay of the $D_{s}^{+}$ meson
into three pions.
Experimental data are available only for the final state
$\pi^{+}\pi^{+}\pi^{-}$ \cite{R1, R2}, but not for the
$\pi^{0}\pi^{0}\pi^{+}$ one.
By inspection of the results, we observe that the quasi two
body state $f_{0}(980) \pi^{+}$ has been clearly identified \cite{R1}
whereas for the quasi
two body $\rho^{0}(770) \pi^{+}$ state,
no significant signal has been found \cite{R1}.
Another remarkable feature of the data is
an important fraction of the decay proceeds
into a non resonant $ \pi^{+}\pi^{+}\pi^{-}$ state \cite{R1},
this fact necessarily
implies a sizable $W^{+}$ annihilation  contribution.
More precisely, the rate for
$D_{s}^{+}$ decays into a non resonant
$\pi^{+}\pi^{+}\pi^{-}$ state is about
1/3 of the dominant spectator decay rate
$D_{s}^{+} \rightarrow \phi \pi^{+} $,
both being governed by the same
Cabibbo-Kobayashi-Maskawa (CKM) favored factor
$V_{cs}^{*} V_{ud}$.

After a brief discussion of the kinematics for three body decay
and a presentation of the experimental results in section II,
we construct a three component model for the transition
amplitude of the mode $D_{s}^{+}\rightarrow \pi^{+}\pi^{+}\pi^{-} $.
The first component corresponds to a quasi two body
$ \rho^{0} \pi^{+} $ state, followed by $\rho^{0} \ra \pi^{+}\pi^{-} $.
Since $\rho^{0}$ meson is a $(u\ol{u} - d\ol{d} )/\sqrt{2} $ isovector state,
the decay $D_{s}^{+} \rightarrow \rho^{0}\pi^{+} $
can proceed only through the $W$ annihilation mechanism (henceforth WA).
In section III, we propose a model for the
$ W^{+} \rightarrow \rho^{0} \pi^{+} $ transition involving
an intermediate virtual $\pi(1300)$ meson.
The experimental upper limit for the branching ratio
$\cB (D_{s}^{+} \rightarrow \rho^{0}\pi^{+} $) allows us
to derive an upper bound for the decay constant $f_{\pi^{'}}$
of the $\pi(1300)$ meson to be around 10 MeV.

The second component is associated to the
observed quasi two body $f_{0} \pi^{+}$ state followed
by the decay $f_{0} \rightarrow \pi^{+}\pi^{-}$.
The quark content of the  $f_{0}(980)$ meson is crucial
for constructing the amplitude
$D_{s}^{+} \rightarrow f_{0} \pi^{+}$.
If the $f_{0}$ is a pure  $(u\ol{u} + d\ol{d} )/\sqrt{2} $ isoscalar state,
we must again use the WA,
as for the $\rho^{0}$ previously.
In section IV we show that by comparing
the $D_{s}^{+}\rightarrow \rho^{0} \pi^{+} $ and
$D_{s}^{+}\rightarrow f_{0} \pi^{+} $ rates, such an assignment
$(u\ol{u} + d\ol{d} )/\sqrt{2} $ for $f_{0}(980)$ is in strong disagreement
with experiment.
Then the $f_{0}(980)$ must content a large  $s\ol{s}$ component or even be
a pure $s\ol{s}$  state.
Now the spectator mechanism can be used for the  decay
$ D_{s}^{+} \rightarrow f_{0} \pi^{+} $,
and from experiment, we estimate in section IV the magnitude of the hadronic
form factor $F_{0}^{D_{s}f_{0}} (m_{\pi}^{2})$ describing
the $ D_{s} \ra f_{0} W^{+} $ transition.
If $f_{0}(980)$ is a pure  $s\ol{s}$ state, we obtain,
from experiment $F_{0}^{D_{s}f_{0}}(m_{\pi}^{2})$ $\approx$ 0.36,
a reasonable value.

For the third component associated to a non resonant
$\pi^{+}\pi^{+}\pi^{-}$ state,
we introduce a complex constant amplitude
which corresponds to an uniformly populated Dalitz plot.
Such a component is determined, from experiment, in section V
where, in addition, the total
rate for $D_{s}^{+} \ra \pi^{+}\pi^{+}\pi^{-}$ is computed
using the three component model described above.
The result, consistent with experimental data,
is presented as a function of a possible phase
for the constant non resonant amplitude.

The $\pi^{+}$ and $\pi^{-}$ energy distributions for the decay
$D_{s}^{+} \ra \pi^{+}\pi^{+}\pi^{-}$ are computed in section VI.
The $f_{0}\pi^{+}$ quasi two body state is clearly seen in the $\pi^{+}$ energy
distribution, as already  obtained in \cite{R1},
no  $\rho^{0}\pi^{+}$ enhancement is present,
and the $\pi^{-}$ energy distribution does not reveal any particular structure
even if it is very different from the pure phase space distribution.
 It is then straightforward to make predictions for the decay mode
$D_{s}^{+}  \ra\ \pi^{0}\pi^{0}\pi^{+}$ using isospin considerations
relating the two final states $\pi^{+}\pi^{+}\pi^{-}$
and $\pi^{0}\pi^{0}\pi^{+}$.
In section VII, we present our results for the rate
and for the $\pi^{0}$, $\pi^{+}$ energy distributions
of the $D_s^{+}$ decay into $\pi^{0}\pi^{0}\pi^{+}$.

Finally, we end this introduction by few considerations concerning
the quark content of the $f_{0}(980)$ scalar meson based firstly
on SU(3) flavor symmetry and secondly
on the decay products of the $f_{0}(980)$.
The SU(3) flavor nonet $J^{PC}$ = $O^{++}$ contents an isovector meson
$a_{0}(980)$, two isoscalar mesons $f_{0}(980)$ and $f_{0}(1300)$
and two isodoublets of strange mesons $K_{0}^{*}(1490)$.
The physical particles $f_{0}(980)$ and $f_{0}(1300)$
are expected, as usual, to be the result of a configuration mixing
between the isoscalar octet and singlet weights.
We remark that the quasi degeneracy in mass between the $a_{0}(980)$
and $f_{0}(980)$ is reminiscent of that between
the $\rho(770)$ and $\omega(782)$
in the $J^{PC}$ = $ 1^{--}$ nonet.
A first naive expectation is that $f_{0}(980)$,
like $\omega(782)$,
is dominantly a $(u\ol{u}+d\ol{d})/\sqrt{2}$ bound state,
the $f_{0}(1300)$ like $\phi(1020)$ being essentially a $s\ol{s}$ state.
However such an assignment is wrong for at least two reasons
from both theoretical and experimental considerations.
Indeed if one tries to apply for the scalar $J^{PC}$ = $O^{++}$ nonet
the usual mixing formalism which has been very successful
in the $J^{PC}$ = $1^{--}$, $0^{-+}$, $2^{++}$, $3^{--}$ cases,
then we are faced with the impossibility to determine the mixing angle
$\theta_{s}$ for scalar mesons. Because of the large mass of
the $K_{0}^*(1490)$ meson, the formalism leads to $tan^{2}\theta_{s} < 0$,
no matter is used the quadratic or linear Gell-Mann - Okubo mass formula.
Therefore the previous analogy between the $J^{PC} = 1^{--}$
and $ 0^{++}$ nonets is meaningless.
 The second reason is an examination of
the experimental decay products of the $f_{0}(980)$
and $f_{0}(1300)$, in particular the $\pi\pi$ and $K \ol{K} $
channels.
In spite of the tiny phase space available
for $f_{0}(980)\ra K\ol{K}$,
its branching ratio  $22 \%$ is actually a very large number.
The situation is inverted for the second meson $f_{0}(1300)$, the available
phase space for $K \ol{K}$ is much larger, however its branching ratio is only
$7.5 \%$, with $93.6 \%$ into $\pi\pi$.
{}From these experimental considerations, we conclude that if the $J^{PC} =
0^{++}$ nonet is made of quark-antiquark bound states, the $f_{0}(980)$ must
contain  a large $s\ol{s}$ component and the $f_{0}(1300)$ a small $s\ol{s}$
one, with a $(u\ol{u}+d\ol{d})/\sqrt{2}$ component being dominant for
$f_{0}(1300)$
and secondary for $f_{0}(980)$.
Moreover, we shall see in section IV that the comparison of the
$D_{s}\ra \rho^{0}\pi^{+}$ and $D_{s}\ra f_{0}(980)\pi^{+}$ rates also
confirms the dominant $s\ol{s}$ component of the $f_{0}(980)$ meson.
Such $s\ol{s}$  classification for $f_{0}(980)$ has also been recently
advocated in \cite{RT}.

\vs{7mm}
\large{} {\bf II \hs{3mm}
Generalities and Experimental results.}
\vs{5mm}
\normalsize

\large
1). \normalsize
We study the decay of the $D_s^{+}$ meson of energy momentum
$ P $ into three $ \pi $ mesons of energy momenta
$ p_{1} $, $ p_{2} $, $ p_{3} $ with the relation
$ P = p_{1} + p_{2} +p_{3} $.
We introduce the Mandelstam variables
$ s_{1}, s_{2}, s_{3} $ and the $ \pi $ meson energies
in the $ D_{s}^+ $ rest frame $ E_{1}, E_{2}, E_{3} $.
At the level of accuracy of the experiments,
it is legitimate to neglect the mass difference between charged
and neutral pions and we get
\beqa
s_{1}  &=& ( p_2 + p_3 )^2 = ( P - p_1 )^2
= m_{D_s}^2 + m_{\pi}^2 - 2 m_{D_s} E_1   \no \\
s_{2} &=&  ( p_3 + p_1 )^2 =  ( P - p_2 )^2
= m_{D_s}^2 + m_{\pi}^2 - 2 m_{D_s} E_2 \label{eq:1} \\
s_{3} &=&   (p_1 +p_2 )^2 = ( P - p_3)^2
= m_{D_s}^2 + m_{\pi}^2 - 2 m_{D_s} E_3 \no
\eeqa
Energy momentum conservation implies the relations
\beq
s_1 + s_2 + s_3 = m_{D_s}^2 + 3 m_{\pi}^2 , \hs{20mm}
E_1 + E_2 + E_3 = m_{D_s} \label{eq:2}
\eeq

The double differential distribution is given
in terms of the transition matrix element by
the expression
\beq
d\Gamma = {{1\over{64 \pi^3}}} \hs{2mm} {{ 1\over{m_{D_s}}}} \hs{2mm}
|< 3\pi | T | D_s^{+}>|^2 \hs{2mm} dE_1 \hs{1mm} dE_2  \label{eq:3}
\eeq
We remind that the transition matrix element
$< 3 \pi | T |D_{s}^{+} >$
involving three body final state is dimensionless.
In the $( E_1, E_2 )$ plane, the phase space is defined by the constraints
\beqa
& &
m_{\pi} \hs{2mm} \leq \hs{2mm} E_1
\hs{2mm} \leq \hs{2mm} {( m_{D_s}^2 - 3 m_{\pi}^2 ) \over {2 m_{D_s} } } \no \\
& &
E_{-}(E_1) \hs{2mm} \leq \hs{2mm} E_2 \hs{2mm} \leq \hs{2mm} E_{+}(E_1)
\label{eq:4}
\eeqa
with
\beq
E_{\pm}(E) = {1 \over 2} \hs{1mm} (m_{D_s} -E ) \pm
\hs{2mm} {1 \over 2} \hs{1mm} \sqrt{ {(E^2-m_{\pi}^2)
(m_{D_s}^2 - 3 m_{\pi}^2 -2m_{Ds}E ) \over m_{Ds}^2 + m_{\pi}^2 - 2m_{Ds} E  }}
\label{eq:5}
\eeq
Of course, the mass difference between charged
and neutral pions being neglected,
we have the same phase space in the two other planes
$(E_2, E_3)$ and $(E_1, E_3)$.
Two possible three pions final states are
$\pi^{+}\pi^{+}\pi^{-}$ and $\pi^{0}\pi^{0}\pi^{+}$.
We make the following choice of $\pi$ meson variables

\begin{center}
(1). \hs{2mm}
final state $\pi^{+}\pi^{+}\pi^{-}$,\hs{18mm}
$E_1(\pi^{+}), \hs{1mm} E_2(\pi^{+}), \hs{1mm} E_3(\pi^{-})$

(2). \hs{2mm}
final state $\pi^{0}\pi^{0}\pi^{+}$,\hs{20mm}
$E_1(\pi^{0}), \hs{1mm} E_2(\pi^{0}), \hs{1mm} E_3(\pi^{+})$
\end{center}

\large
2). \normalsize
Let us now discuss the experimental situation concerning
the only observed 3 pion final state $\pi^{+} \pi^{+}\pi^{-}$.
The decay rate for the mode $D_{s}^{+} \ra  \pi^{+}\pi^{+}\pi^{-}$
has been first measured by the E 691 collaboration at Fermi Lab. \cite{R1}
and the result has been confirmed few years later within error
by the NA 82 collaboration at CERN \cite{R2}.
In addition, the E691 group has been able to exhibit in the
$\pi^{+}\pi^{+}\pi^{-}$ Dalitz plot a structure
due to the $f_{0}(980) \pi^{+}$ quasi two body state.
However a structure due to the $\rho^{0}(770) \pi^{+}$ quasi two body state
has not been clearly identified.
The various branching ratios in \cite{R1} are normalized to
the most easily measured mode $D_{s}^{+} \ra \phi \pi^{+} $.
Using the Particle Data Group value \cite{R3}
\beq
\cB(D_{s}^{+} \ra \phi + \pi^{+}  )  =   (3.5 \pm 0.4 ) \%
\label{eq:6} \\
\eeq
the branching ratios obtained in \cite{R1} are :
\beqa
\cB(D_{s}^{+} \ra  \pi^{+} \pi^{+}\pi^{-} )_{TOT}
&=& (1.54 \pm 0.42) \%
\label{eq:7} \\
\cB(D_{s}^{+} \ra  \pi^{+} \pi^{+}\pi^{-} )_{NR}
&=& (1.015 \pm 0.352) \%
\label{eq:8} \\
\cB(D_{s}^{+} \ra  f_0 \pi^{+} )
&=& (0.98 \pm 0.38) \%
\label{eq:9} \\
\cB(D_{s}^{+} \ra  \rho^0 \pi^{+} )
&<& 0.28 \% \hs{5mm} at \hs{2mm}90 \hs{2mm} \% \hs{2mm} C.L.
\label{eq:10}
\eeqa
where, in (\ref{eq:8}) the subscript $NR$ means non resonant \cite{R1}.
Our results (\ref{eq:8}), (\ref{eq:9}), and (\ref{eq:10}) agree with
those given in \cite{R3}.
However the branching ratio of Eq.(\ref{eq:7}) quoted in \cite{R3}
is slightly smaller, $(1.35 \pm 0.31) \%$,
because an average has been made between the data obtained in \cite{R1}
and \cite{R2}.
For a question of consistency we shall use, in this paper,
the set of data (\ref{eq:7}) $-$ (\ref{eq:10})
in order to preserve the relative fractions of the various terms
as quoted in \cite{R1}.

\newpage
\vs{7mm}
\hs{3mm} \large{ } {\bf III. \hs{3mm} The decay mode
for $D_s^{+} \ra \rho^0  \pi^{+} \ra \pi^{+} \pi^{+} \pi^{-} $ }
\normalsize
\vs{5mm}

\large
1). \normalsize
Let us consider the quasi two body final state $\rho^0\pi$.
The transition matrix element has the form
\beq
< \hs{2mm} \rho^0 \pi^{+} \hs{2mm} | \hs{2mm} T \hs{2mm}| \hs{2mm} D_s^{+} >
\hs{2mm} = \hs{2mm}
g_{D_s\rho \pi} \hs{2mm} (P_{D_s} + p_{\pi})^{\mu} \hs{2mm}
e^{*}_{\mu}(p_{\rho})
\label{eq:11}
\eeq
where $e_{\mu}(p_{\rho})$ describes the $\rho$ meson polarization.
The dimensionless quantity $g_{D_s\rho\pi}$ is a weak complex coupling
constant in terms of which, the decay width is given by
\beq
\G(D_s^{+} \ra \rho^0 \pi^{+} ) \hs{2mm} = \hs{2mm}
{1 \over 2 \pi} \hs{2mm}
{K_{\rho}^3 \over m_{\rho}^2 } \hs{2mm}
|g_{D_s\rho\pi}|^2.
\label{eq:12}
\eeq
Here $K_{\rho}$ is the center of mass momentum of the final $\rho\pi$
 particles in the $D_s^{+}$ meson rest frame.

\large
2). \normalsize
The $\rho^0$ and pion  are  quark-antiquark bound states
containing only
$u$ and $d$ quarks and the decay
$D_s^{+} \ra \rho^{0} \pi^{+}$
can only proceed through the WA,
since both charm  and strange quarks being absent in the final state.
For the transition $W^+ \ra \rho^0 \pi^+$ we assume,
by the Partial Conservation of the Axial Current (PCAC)
the contribution of an intermediate state having
the quantum number of a $\pi^{+}$ meson \cite{R4, R5, R6}
and which might be the $\pi$ meson itself or its recurrence $\pi(1300)$.
It has been checked that the $\pi$ meson intermediate state gives
a contribution many order of magnitude smaller than that of the $\pi(1300)$
and only the latter one is retained in the diagram of Fig.1.

Here $\pi^{'} \equiv \pi(1300)$ and the corresponding expression of
the weak decay constant $g_{D_s\rho\pi}$ is given by
\beq
g_{D_s\rho\pi} \hs{2mm} = \hs{2mm}
a_1 \hs{2mm} {G_F \hs{1mm} m_{D_s}^2 \over \sqrt{2} } \hs{2mm}
V^{*}_{cs} \hs{1mm} V_{ud} \hs{2mm}
{f_{D_s} \hs{2mm} f_{\pi^{'}} \over m_{\pi^{'}}^2 - m_{D_s}^2 - i \hs{1mm}
m_{\pi^{'}} \hs{1mm} \G_{\pi^{'}}(m_{D_s}^2) } \hs{2mm}
g_{\pi^{'} \rho \pi}
\label{eq:13}
\eeq
where $a_1$ is the phenomenological parameter introduced by
Bauer, Stech, Wirbel (henceforth BSW) \cite{R7};
$V_{cs}, V_{ud}$ the relevant CKM matrix elements;
$f_{D_s}$ and $f_{\pi^{'}}$ are the leptonic decay constants of the $D_s^{+}$
and $\pi^{'}$ mesons respectively.

The dimensionless strong decay constant
$g_{\pi^{'}\rho\pi}$ is related to the corresponding decay rate
by
\beq
\G(\pi^{'} \ra \rho^0 \pi^{+}) \hs{2mm} = \hs{2mm}
{1 \over 2 \pi} \hs{2mm}
{{K_{\rho}^{'}}^3 \over m_{\rho}^2 } \hs{2mm}
|g_{\pi^{'}\rho\pi}|^2.
\label{eq:14}
\eeq
where $K_{\rho}^{'}$ is the C. M. momentum in the $\pi^{'}$ rest frame
of the $\rho \pi$ final state.

Combining Eqs. (\ref{eq:12}), (\ref{eq:13}) and (\ref{eq:14}) we obtain
\beq
\cB(D_s^{+} \ra \rho^0 \pi^{+} ) \hs{2mm} = \hs{2mm}
{\tau_{D_s} \over \hbar } \hs{2mm}
\left( {K_{\rho} \over K_{\rho}^{'}} \right)^{3} \hs{2mm}
f_{\pi^{'}}^2 \hs{1mm} \Lambda^2 \hs{1mm}
\cB(\pi^{'} \ra \rho^{0} \pi^{+}) \label{eq:15}
\eeq
where $\Lambda^2$ is given by
\beq
\Lambda^2 \hs{2mm} = \hs{2mm}
a_1^2 \hs{2mm} \left[ {G_F \hs{1mm} m_{D_s}^2 \over \sqrt{2} } \right]^2
\hs{2mm} |V_{cs}|^2 \hs{2mm} |V_{ud}|^2 \hs{2mm}
f_{D_s}^2 \hs{2mm}
{ \G_{\pi^{'}}(m_{\pi^{'}}^2) \over
(m_{\pi^{'}}^2 - m_{D_s}^2 )^2 + m_{\pi^{'}}^2 \hs{1mm}
\G_{\pi^{'}}^2(m_{D_s}^2)}
\label{eq:16}
\eeq

\large
3). \normalsize
For masses, widths and CKM matrix elements
we use the values collected by the Particle Data Group \cite{R3}.
The quantity $f_{D_s}$ is choosen to be
$f_{D_s} = 280 \hs{2mm} MeV$ consistent with recent experimental data
\cite{R3} and theoretical expectations.
For the decay $\pi(1300) \ra \rho \pi$ we use \cite{R3}
\beq
\cB(\pi^{'} \ra \rho  \pi) \hs{2mm} = \hs{2mm} 0.3205
\label{eq:17}
\eeq
and we leave $f_{\pi^{'}}$ as a free parameter.
For $a_1$ we take $a_1 = 1.26$ \cite{R8}.

In the following, we use the simple energy dependence
$m \G(s) = \sqrt{s} \hs{1mm} \G(m^2)$ for the widths,
although more sophisticated
expressions have been proposed \cite{R4, R5}.
Because of the not too large difference between the $\pi^{'}$
and $D_s^{+}$ masses,
we expect the sensitivity of
Eqs.(\ref{eq:13}) and (\ref{eq:16}) to different forms of
$\G_{\pi^{'}}(m_{D_s}^2)$
to be relatively modest.

We now write the branching ratio $\cB(D_s^+ \ra \rho^0 \pi^+)$ in the form
\beq
\cB(D_s^{+} \ra \rho^0 \pi^{+}) \hs{2mm} = \hs{2mm}
C^2 \hs{2mm} f_{\pi^{'}}^2
\label{eq:18}
\eeq
We retain only in the numerical computation of $C^2$ the large
uncertainty due to the $\pi(1300)$ width \cite{R3},
$\G_{\pi^{'}} = (400 \pm 200) \hs{2mm} MeV$ and we get
\beq
C^2 \hs{2mm} = \hs{2mm} 16.3 \hs{2mm}^{+ \hs{2mm} 5.1}_{- \hs{2mm} 7.4}
\hs{3mm} GeV^{-2}
\label{eq:19}
\eeq
Using now the $90 \%$ confidence level upper limit
$\cB(D_s \ra \rho^0 \pi^{+}) < 0.0028$ of Eq.(\ref{eq:10})
we deduce an upper limit for $f_{\pi^{'}}$
\beq
\left| \hs{2mm} f_{\pi^{'}} \hs{2mm} \right| \hs{2mm} < \hs{2mm}
\left| \begin{array}{c} 11.45\\ 13.11 \\ 17.72  \end{array}
\right| \hs{2mm} MeV
\hs{4mm} for \hs{4mm}
\G_{\pi^{'}} \hs{2mm} = \hs{2mm}
\left| \begin{array}{c} 600 \\ 400 \\ 200  \end{array}
\right| \hs{2mm} MeV
\label{eq:20}
\eeq
Theoretical value for $f_{\pi^{'}}$, as large as $40 \hs{2mm} MeV$,
seems to be overestimated in the literature \cite{R6}.
Eq.(\ref{eq:20}) gives, to our knowledge,
the first experimental information on $f_{\pi^{'}}$,
and our result is compatible with
theoretical calculations using non-relativistic chiral quark model \cite{R9}
or $QCD$ sum rule technique \cite{R10}.

\large
4). \normalsize
We now take into account the non-zero $\rho$ meson width effect.
The production of the quasi two body state $\rho^0 \pi^{+}$ is followed
by the decay $\rho^0 \ra \pi^+ \pi^-$.
For the final state $\pi^+\pi^-\pi^+$,
$E_1$ and $E_2$ are the $\pi^{+}$ energies and $E_3$ is the $\pi^{-}$ energy.
The decay amplitude for $D_s^+ \ra \rho^0 \pi^{+} \ra \pi^+\pi^-\pi^+$
is given by
\beq
<\pi^+\pi^-\pi^+ \hs{1mm} | \hs{1mm} T \hs{1mm}| \hs{1mm} D_s^+ >_{\rho\pi}
\hs{2mm} = \hs{2mm}
g_{D_s\rho\pi} \hs{1mm} g_{\rho\pi\pi} \hs{1mm}
\left\{ {s_2 - s_3 \over
m_{\rho}^2 - s_1 - i \hs{1mm} m_{\rho} \hs{1mm}\G_{\rho}(s_1)}
\hs{2mm} + \hs{2mm}
{s_1 - s_3 \over
m_{\rho}^2 - s_2 - i \hs{1mm} m_{\rho} \hs{1mm}\G_{\rho}(s_2)}
\right\}
\label{eq:21}
\eeq
and the corresponding decay rate has the form
\beq
\G(D_s^{+} \ra \rho^0 \pi^+ \ra \pi^+ \pi^- \pi^+) \hs{2mm} = \hs{2mm}
{1 \over 64 \hs{1mm} \pi^3} \hs{2mm} m_{D_s} \hs{2mm}
|g_{D_s\rho \pi}|^2 \hs{2mm} |g_{\rho \pi \pi}|^2 \hs{2mm}
{J \over 2}
\label{eq:22}
\eeq
where $J$ is the phase space integral
\beq
J \hs{2mm} = \hs{2mm}
{1 \over m_{D_s}^2} \int \int \hs{1mm} dE_1 \hs{1mm} dE_2 \hs{2mm}
\left| {s_2 - s_3 \over
m_{\rho}^2 - s_1 - i \hs{1mm} \sqrt{s_1} \hs{1mm}\G_{\rho}}
\hs{2mm} + \hs{2mm}
{s_1 - s_3 \over
m_{\rho}^2 - s_2 - i \hs{1mm} \sqrt{s_2} \hs{1mm}\G_{\rho}}
\right|^2
\label{eq:23}
\eeq
The energy dependence of the $\rho$ meson width has been taken as
explained previously and the factor ${1 \over 2}$ in Eq.(\ref{eq:22})
is due to the presence of two identical $\pi^{+}$ in the final state.

The dimensionless constant $g_{\rho\pi\pi}$ is related to
the $\rho$ meson width by
\beq
\G(\rho^0 \ra \pi^+\pi^-) \hs{2mm} = \hs{2mm}
{1 \over 6 \pi} \hs{2mm} {K_{\pi}^3 \over m_{\rho}^2} \hs{2mm}
(g_{\rho\pi\pi})^2
\label{eq:24}
\eeq
where $K_{\pi}$ is the C. M. momentum in the $\rho^0$ meson rest frame
of the two pions.

Combining Eq.(\ref{eq:22}) and (\ref{eq:24}) we obtain
\beq
\G(D_s^+ \ra \rho^0\pi^+ \ra \pi^+\pi^-\pi^+) \hs{2mm} = \hs{2mm}
{3 \over 8 \hs{1mm} \pi^2} \hs{2mm} {m_{D_s} \over m_{\rho}} \hs{2mm}
|g_{D_s\rho \pi}|^2 \hs{2mm}
{\G(\rho^0 \ra \pi^+\pi^-) \over (1 - 4 m_{\pi}^2 / m_{\rho}^2)^{3/2}}
\hs{2mm} J
\label{eq:25}
\eeq
It is straightforward to estimate the importance of
the non-zero $\rho^0$ meson width.
The result is obviously independent of $g_{D_s\rho\pi}$ and
comparing Eq.(\ref{eq:12}) and (\ref{eq:22}) we get
\beq
{\G(D_s^+ \ra \rho^0 \pi^+ \ra \pi^+\pi^-\pi^+)
\over \G(D_s^+ \ra \rho^0 \pi^+) }
\hs{2mm} = \hs{2mm}
{3 \over 4 \pi} \hs{2mm}
{m_{\rho} m_{D_s} \over K_{\rho}^3} \hs{2mm}
{\G(\rho^0 \ra \pi^+\pi^-) \over (1 - 4 m_{\pi}^2 / m_{\rho}^2)^{3/2}}
\hs{2mm} J
\label{eq:26}
\eeq
The value of $ J $ obtained by numerical integration is $J = 8.0997$
and we get
\beq
{\G(D_s^+ \ra \rho^0 \pi^+ \ra \pi^+\pi^-\pi^+)
\over \G(D_s^+ \ra \rho^0 \pi^+) }
\hs{2mm} = \hs{2mm}
0.9719 \pm 0.0096
\label{eq:27}
\eeq
where the error is due to the uncertainty of the $\rho^0$ meson width.

Finite $\rho^0$ meson width effect is small and the upper limits
obtained in Eq. ({\ref{eq:20}) for $|f_{\pi^{'}}|$
are only slightly modified
\beq
\left| \hs{2mm} f_{\pi^{'}} \hs{2mm} \right| \hs{2mm} < \hs{2mm}
\left| \begin{array}{c} 11.61\\ 13.30 \\ 17.98  \end{array}
\right| \hs{2mm} MeV
\hs{4mm} for \hs{4mm}
\G_{\pi^{'}} \hs{2mm} = \hs{2mm}
\left| \begin{array}{c} 600 \\ 400 \\ 200  \end{array}
\right| \hs{2mm} MeV
\label{eq:28}
\eeq

\vs{7mm}
\hs{3mm} \large{ } {\bf IV. \hs{3mm} The decay mode
for $D_s^{+} \ra f_0 \pi^{+} \ra \pi^{+} \pi^{+} \pi^{-} $ }
\normalsize
\vs{5mm}

\large
1). \normalsize
We consider now the quasi two body final state $f_0(980) \pi^{+}$.
Since two body decay amplitude has
 a mass dimension, let us write the transition matrix element
for spinless particles in the form,
\beq
< f_0 \hs{2mm} \pi^{+} \hs{2mm}| \hs{2mm} T \hs{2mm} | \hs{2mm} D_s^{+} >
\hs{2mm}
\equiv \hs{2mm} m_{D_s} \hs{2mm} g_{D_s f_0 \pi}
\label{eq:29}
\eeq
where the dimensionless quantity $g_{D_s f_0 \pi} $ is a weak complex constant
in terms of which, the decay width is given by
\beq
\G( \hs{2mm} D_s^{+} \ra f_0 \hs{2mm} \pi^{+}) = {1 \over 8 \pi} \hs{2mm}
K_f \hs{2mm} \left|g_{D_s f_0 \pi} \right|^2 .     \label{eq:30}
\eeq
Here $K_f$ is the C. M. momentum of the final particles $f_0 \pi^{+}$
in the $D_s^{+}$ rest frame.

\large
2). \normalsize
The considerations made in the introduction concerning the quark content of
the $f_0(980)$ meson can be tested in the decay of interest here,
$D_s^{+} \ra f_0 \pi^{+}$.
Let us first assume that $f_0(980)$ is a
$(u\ol{u} + d\ol{d})/\sqrt{2}$ bound state,
then we can show that this assumption is in contradiction with
 experimental data.
Indeed under such assumption, the only possible mechanism
for both decays $D_s \ra f_0 \pi^{+}$
and $D_s \ra \rho^0 \pi^{+}$ is the WA and the coupling constant
$g_{D_s f_0 \pi} $ is obtained from a diagram similar to the one of Fig.1
where the $\rho^0$ is replaced by the $f_0$.
It is straightforward to derive a relation
between the two weak coupling constants
\beq
\frac{(g_{D_s f_0 \pi})_{WA}}{g_{D_s \rho \pi} } \
 = {m_{\pi^{'}} \over m_{D_s}} \hs{2mm}
{g_{\pi^{'} f_0 \pi} \over  g_{\pi^{'} \rho \pi} } \label{eq:31}
\eeq
where the subscript $WA$ indicates that the decay amplitude
is performed in the $W$ annihilation model of Fig.1,
from which we get
\beq
{\G(D_s^{+} \ra f_0 \pi^{+})_{WA} \over \G(D_s^{+} \ra \rho^0 \pi^{+}) }
\hs{2mm} = \hs{2mm}
\left( { K_{\rho}^{'} \over K_{\rho}} \right)^3 \hs{2mm}
\left( { K_{f} \over K_f^{'}} \right) \hs{2mm}
\left( {m_{\pi}^{'} \over m_{D_s} } \right)^2 \hs{2mm}
{\G(\pi^{'} \ra f_0 \pi^{+}) \over \G(\pi^{'} \ra \rho^0 \pi^{+})}
\label{eq:32}
\eeq
where $K_f^{'}$ is the C. M. momentum in the $\pi^{'}$ rest frame
associated to the decay $\pi^{'} \ra f_0 \pi^{+}$.

We assume that the experimental result of the $\pi^{'}$ decay given in
\cite{R3} for the final state $(\pi\pi)_{S-wave} + \pi^{+}$
corresponds to $f_0 \pi^{+}$ and we use
\beq
{\G(\pi^{'} \ra f_0 \pi^{+}) \over \G(\pi^{'} \ra \rho^0 \pi^{+})}
\hs{2mm} = \hs{2mm} 2.12 . \label{eq:33}
\eeq

Numerically, we obtain from Eq.({\ref{eq:32})
\beq
\G(D_s^{+} \ra f_0 \pi^{+})_{WA} \hs{2mm} = \hs{2mm}
0.318 \hs{1mm} \cdot \hs{1mm} \G(D_s^{+} \ra \rho^0 \pi^{+}). \label{eq:34}
\eeq
Using the upper limit (\ref{eq:10}) for the rate of
$D_s^{+} \ra \rho^0 \pi^{+}$, we get the prediction
\beq
\cB(D_s^{+} \ra f_0 \pi^{+})_{WA} \hs{2mm} < \hs{2mm}
8.9 \hs{1mm} \cdot \hs{1mm} 10^{-4} \label{eq:35}
\eeq
in complete disagreement with the observed experimental value
Eq.(\ref{eq:9}).

It follows that the dominant component of $f_0(980)$
cannot be  $(u\ol{u} + d\ol{d})/\sqrt{2}$ and a large $s\ol{s}$
component must be present.
In fact such an $s\ol{s}$ component will be responsible
for the observed strong decay mode $f_0 \ra K\ol{K}$.

\large
3). \normalsize
Having shown that $f_0(980)$ is mainly a pure $s\ol{s}$ state
by both considerations based firstly on its decay modes into $K\ol{K}$,
secondly on those of $D_s^+$ into $f_0 \pi$
(beside the mixing in the $J^{PC} = 0^{++}$ $SU(3)$ flavor
nonet argument mentioned previously in the Introduction),
the dominant decay mechanism for the mode $D_s^{+} \ra f_0 \pi^{+}$
is then the spectator diagram.
The weak coupling constant $g_{D_s f_0 \pi}$
is now computed from the diagram of Fig.2 and the result is
\beq
g_{D_s f_0 \pi} \hs{2mm} = \hs{2mm} i \hs{2mm}
a_1 \hs{2mm} \frac{G_F m_{D_s}^2}{\sqrt{2}} \hs{2mm}
V_{cs}^{*} \hs{2mm} V_{ud} \hs{2mm}
{(m_{D_s}^2 - m_{f_0}^2) \over m_{D_s}^2} \hs{2mm}
{f_{\pi} \over m_{D_s}} \hs{2mm}
F_{0}^{D_sf_0}(m_{\pi}^2) \label{eq:36}
\eeq
where $F_0^{D_sf_0}(q^2)$ is the $D_s \ra f_0$ hadronic form factor
in the BSW notation
associated to the spin zero part of the weak
axial vector current.
Using the experimental branching ratio
$\cB(D_s^{+} \ra f_0 \pi^{+})$ in Eq. (\ref{eq:9}),
we deduce the value of the hadronic form factor
$F_0^{D_s f_0}(m_{\pi}^2)$
\beq
\left| F_0^{D_sf_0}(m_{\pi}^2) \right|
\hs{2mm} = \hs{2mm}
0.36 \hs{2mm}^{+ \hs{2mm} 0.06}_{- \hs{2mm} 0.08} \label{eq:37}
\eeq
which appears to be an acceptable order of magnitude
for an hadronic form factor.

If the $f_0(980)$ is not a pure $s\ol{s}$ state,
what has been obtained numerically in Eq.(\ref{eq:37})
is the product of $F_0^{D_sf_0}(m_{\pi}^2)$ by a mixing coefficient
$\mu$ corresponding to the amount of $s\ol{s}$ contained in $f_0(980)$.
Of course, $| \mu| \leq 1 $, such that the true form factor is
$|F_0^{D_sf_0}(m_{\pi}^2)| \geq 0.36^{+ \hs{2mm} 0.06}_{- \hs{2mm} 0.08}$.

It is interesting to compare this phenomenological value with
the similar one coming from the well measured
$D_s^+ \ra \phi \pi^+$ decay mode.
In the ideal mixing situation close to reality,
the $\phi$ meson is a pure
$s\ol{s}$ state and the decay $D_s^+ \ra \phi \pi^+$ is described by
the spectator mechanism involving the hadronic form factor
$A_0^{D_s\phi}(m_{\pi}^2)$.
Using the experimental result quoted in Eq.({\ref{eq:6}) we obtain
an experimental value for $A_0^{D_s\phi}(m_{\pi}^2)$
\beq
| A_0^{D_s\phi}(m_{\pi}^2) | \hs{2mm} = \hs{2mm}
0.72 \hs{2mm} \pm \hs{2mm} 0.04
\label{eq:A0}
\eeq
Of course this form factor is larger than the one previously obtained for
$\mu \hs{1mm} F_0^{D_sf_0}(m_{\pi}^2)$, but there are no reasons why
they should be equal.

However let us remind that both
$F_0^{D_sf_0}$ and $A_0^{D_s\phi}$ are determined by
the factorization method.
It is possible that non factorization contributions will change
these numerical estimates in a sizable way,
unfortunately, to our knowledge, a convincing approach
to compute the non factorization term is not yet available.

\large
4). \normalsize
We now take into account the $f_0$ meson width effect.
The decay amplitude for $D_s^{+} \ra f_0 \pi^{+} \ra \pi^{+}\pi^{-}\pi^{+}$
is obtained with the help of the definition
Eq.(\ref{eq:29}), together with the similar expression
associated to the decay $f_0 \ra \pi^{+}\pi^{-}$.
\beqa
< \pi^{+}\pi^{-}\pi^{+} \hs{2mm} | \hs{2mm} T \hs{2mm} | \hs{2mm} D_s^{+}
>_{f_0\pi} \hs{2mm} &=& \hs{2mm}
g_{D_sf_0\pi} \hs{1mm} g_{f_0\pi^{+}\pi^{-}} \hs{2mm}
{m_{D_s} \over m_{f_0}} \hs{2mm} \no \\
& &
\left\{ {m_{f_0}^2 \over m_{f_0}^2 - s_1 - i \hs{2mm} m_{f_0} \G_{f_0}(s_1)}
+
{m_{f_0}^2 \over m_{f_0}^2 - s_2 - i \hs{2mm} m_{f_0} \G_{f_0}(s_2)}
\right\} \label{eq:38}
\eeqa
and the corresponding decay rate is
\beq
\G(D_s^{+} \ra f_0 \pi^{+} \ra \pi^{+} \pi^{-} \pi^{+} )
\hs{2mm} = \hs{2mm}
{1 \over 64 \hs{1mm} \pi^3} \hs{2mm}
{m_{D_s}^3 \over m_{f_0}^2} \hs{2mm} |g_{D_sf_0\pi}|^2 \hs{2mm}
g_{f_0\pi^{+}\pi^{-}}^2 \hs{2mm}
{K \over 2}
\label{eq:39}
\eeq
where the dimensionless coefficient $K$ is the phase space integral
\beq
K \hs{2mm} = \hs{2mm}
{m_{f_0}^4 \over m_{D_s}^2} \hs{2mm}
\int \int \hs{1mm} dE_1 \hs{1mm} dE_2 \hs{1mm}
\left| {1 \over m_{f_0}^2 - s_1 - i \hs{2mm} \sqrt{s_1} \hs{1mm} \G_{f_0}}
+ {1 \over m_{f_0}^2 - s_2 - i \hs{2mm} \sqrt{s_2} \hs{1mm} \G_{f_0}}
\right|^2 \label{eq:40}
\eeq

The energy dependence of the $f_0$ meson width
has been taken as explained previously and the factor
${1 \over 2}$ in Eq.(\ref{eq:39}) is due to
the presence of two $\pi^{+}$ in the final state.

The dimensionless constant $g_{f_0\pi^{+}\pi^{-}}$ is related
to the partial decay width by
\beq
\G(f_0 \ra \pi^{+} \pi^{-})
\hs{2mm} = \hs{2mm}
{1 \over 8 \hs{1mm} \pi } \hs{2mm}
K_{\pi} \hs{2mm} ( \hs{2mm} g_{f_0\pi^{+}\pi^{-}} \hs{2mm} )^2
\label{eq:41}
\eeq
where $K_{\pi}$ is the C.M. momentum of the two pions in the $f_0$
rest frame.

Combining Eqs.(\ref{eq:39}) and (\ref{eq:41}) we obtain
\beq
\G(D_s^{+} \ra f_0 \pi^{+} \ra \pi^{+} \pi^{-} \pi^{+} )
\hs{2mm} = \hs{2mm}
{1 \over 8 \hs{1mm} \pi^2} \hs{2mm}
\left( {m_{D_s} \over m_{f_0}} \right)^3 \hs{2mm}
{\G(f_0 \ra \pi^{+} \pi^{-}) \over \sqrt{1 - 4 m_{\pi}^2/m_{f_0}^2} }
\hs{2mm} K \hs{2mm}
\left| g_{D_sf_0\pi} \right|^2 \label{eq:42}
\eeq
The effect of the non zero $f_0$ width  is estimated
by comparing Eqs.(\ref{eq:30}) and (\ref{eq:42})
\beq
{\G(D_s^{+} \ra f_0 \pi^{+} \ra \pi^{+}\pi^{-}\pi^{+}) \over
\G(D_s^{+} \ra f_0 \pi^{+}) \times \cB(f_0 \ra \pi^{+} \pi^{-}) }
\hs{2mm} = \hs{2mm}
{1 \over \pi} \hs{2mm}
\left( {m_{D_s} \over m_{f_0} } \right)^3 \hs{2mm}
{\G_{f_0}  \over \sqrt{1 - 4 m_{\pi}^2/m_{f_0}^2} } \hs{2mm}
{K \over K_f} \label{eq:43}
\eeq
The value of $K$  obtained by numerical integration is
$K = 5.7217$.
We get
\beq
{\G(D_s^{+} \ra f_0 \pi^{+} \ra \pi^{+}\pi^{-}\pi^{+}) \over
\G(D_s^{+} \ra f_0 \pi^{+}) \times \cB(f_0 \ra \pi^{+} \pi^{-}) }
\hs{2mm} = \hs{2mm} 0.9886 \pm 0.1893
\label{eq:44}
\eeq
where the error is due to the uncertainty on the $f_0$ meson width
taken to be $\G_{f_0} = 47 \pm 9 $ $MeV$.

The $f_0$ finite width effect is small and the value of
$F_0^{D_sf_0}(m_{\pi^{'}}^2) $
obtained in Eq.(\ref{eq:37}) is only slightly modifed.

\vs{7mm}
\hs{3mm} \large{ } {\bf V. \hs{3mm} Total rate
for the decay $D_s^{+} \ra \pi^{+} \pi^{+} \pi^{-} $ }
\normalsize
\vs{5mm}

\large
1). \normalsize
We propose a model where the decay amplitude
$< \pi^{+}\pi^{-}\pi^{+} \hs{1mm} | \hs{1mm} T \hs{1mm} | \hs{1mm} D_s^{+}>$
is written as a sum of three contributions
\beqa
< \pi^{+}\pi^{-}\pi^{+} \hs{1mm} | \hs{1mm} T \hs{1mm} | \hs{1mm} D_s^{+}>
\hs{2mm} = \hs{2mm}
< \pi^{+}\pi^{-}\pi^{+} \hs{1mm} | \hs{1mm} T \hs{1mm} | \hs{1mm}
D_s^{+}>_{\rho \pi} \hs{1mm} & + & \hs{1mm}
< \pi^{+}\pi^{-}\pi^{+} \hs{1mm} | \hs{1mm} T \hs{1mm} | \hs{1mm}
D_s^{+}>_{f_0\pi}  \no \\
\hs{1mm} & + & \hs{1mm}
< \pi^{+}\pi^{-}\pi^{+} \hs{1mm} | \hs{1mm} T \hs{1mm} | \hs{1mm} D_s^{+}>_{NR}
\label{eq:45}
\eeqa
The two first components in Eq.(\ref{eq:45}) are associated
to the quasi two body final states $\rho^{0}\pi^{+}$ and $f_0\pi^{+}$
and their expressions have been given in Eqs.(\ref{eq:21}) and (\ref{eq:38}).
The third component describes a non resonant $\pi^{+}\pi^{-}\pi^{+}$ state
and  is naturally assumed to proceed through the $W^{+}$
annihilation mechanism.
In this case the general structure of the decay matrix element is
\beq
< \pi^{+}\pi^{-}\pi^{+} \hs{1mm} | \hs{1mm} T \hs{1mm} | \hs{1mm} D_s^{+}>_{WA}
\hs{2mm} = \hs{2mm} i \hs{2mm}
a_1 \hs{1mm} {G_F \over \sqrt{2}} \hs{1mm}
V_{cs}^{*} \hs{1mm} V_{ud} \hs{1mm} f_{D_s} \hs{1mm} P^{\mu}_{D_s} \hs{1mm}
< \pi^{+}\pi^{-}\pi^{+} \hs{1mm} | \hs{1mm} \cA_{\mu} \hs{1mm} | \hs{1mm} 0 >
\label{eq:46}
\eeq
The matrix element  of the divergence of the weak axial vector current between
the vacuum and the three pion final state is an unknown structure function
$F(E_1, E_2)$ depending on two independent variables, choosen as
the pion energies.
\beq
P^{\mu}_{D_s} \hs{1mm}
< \pi^{+}\pi^{-}\pi^{+} \hs{1mm} | \hs{1mm} \cA_{\mu} \hs{1mm} | \hs{1mm} 0 >
\hs{2mm} = \hs{2mm}
m_{D_s} \hs{1mm} F(E_1, E_2)
\label{eq:47}
\eeq
Since the matrix elememt of the three body decay is dimensionless,
Eq.(\ref{eq:46}) implies that the function $F(E_1,E_2)$
is also dimensionless and is related to the function $F_4$ of \cite{R4}
by $F(E_1, E_2) = m_{D_s} \hs{1mm} F_4(s_1, s_2, Q^2 = m_{D_s}^2)$.

Combining Eqs.(\ref{eq:46}) and (\ref{eq:47}), we get
\beq
< \pi^{+}\pi^{-}\pi^{+} \hs{1mm} | \hs{1mm} T \hs{1mm} | \hs{1mm} D_s^{+}>_{WA}
\hs{2mm} = \hs{2mm} i \hs{2mm}
a_1 \hs{1mm} {G_F \hs{1mm} m_{D_s}^2 \over \sqrt{2}} \hs{1mm}
V_{cs}^{*} \hs{1mm} V_{ud} \hs{1mm} {f_{D_s} \over m_{D_s} }
\hs{1mm} F(E_1, E_2)
\label{eq:48}
\eeq
The corresponding branching ratio is written in the form
\beq
\cB(D_s^{+} \ra \pi^{+} \pi^{-} \pi^{+} )_{WA} = \cN \times I
\label{eq:49}
\eeq
where $\cN$ is a normalization factor,
\beq
\cN = { \tau_{D_s} \over \hbar } \hs{1mm}
{m_{D_s} \over 64 \hs{1mm} \pi^3} \hs{1mm} {1 \over 2} \hs{1mm}
a_1^2 \left( {G_F \hs{1mm} m_{D_s}^2 \over \sqrt{2} } \right) \hs{1mm}
|V_{cs}|^2 \hs{1mm} |V_{ud}|^2 \hs{1mm}
{f_{D_s}^2 \over m_{D_s}^2}
\label{eq:50}
\eeq
and $I$ a phase space integral given by
\beq
I \hs{2mm} = \hs{2mm}
{1 \over m_{D_s}^2 } \hs{1mm} \int \int
\hs{1mm} dE_1 \hs{1mm} dE_2 \hs{2mm} |F(E_1, E_2)|^2
\label{eq:51}
\eeq
In Eq.(\ref{eq:50}) the factor ${1 \over 2}$ is again due to
 the presence of the two $\pi^{+}$ in the final state.

The value of $\cN$ is computed using the same parameters as in section III
and the result is $\cN = 1.0399 \cdot 10^{-2}$.

\large
2). \normalsize
In the non resonant case we assume the structure function $F$ to be
a constant independent
of the pion energies such as $F(E_1, E_2) = F_{NR}$.
The experimental non resonant rate quoted in Eq.(\ref{eq:8})
allows us to extract the modulus of $F_{NR}$.
The corresponding phase space integral $I_{NR}$ in Eq.(\ref{eq:51})
with constant $F_{NR}$  has the value
\beq
I_{NR} \hs{2mm} = \hs{2mm}
0.1053 \hs{2mm} |F_{NR}|^2
\label{eq:52}
\eeq
and we obtain by using Eqs.(\ref{eq:8}), (\ref{eq:49}) and (\ref{eq:52})
\beq
|F_{NR}| \hs{2mm} = \hs{2mm}
3.04 \hs{2mm}^{+ \hs{2mm} 0.49}_{- \hs{2mm} 0.58}
\label{eq:53}
\eeq

\large
3). \normalsize
A  decomposition formally similar to Eq.(\ref{eq:48})
in terms of $F(E_1, E_2)$ can be conveniently written for the two other
components of Eq.(\ref{eq:45}).
This is obvious for the $\rho\pi$ component which is
of the $W$ annihilation type and we get from Eqs.(\ref{eq:13})
and (\ref{eq:21})
\beq
F_{\rho\pi}(E_1, E_2) \hs{2mm} = \hs{2mm}
A_{\rho\pi} \hs{1mm}\cdot H_{\rho\pi}(E_1, E_2)
\label{eq:54}
\eeq
with
\beq
A_{\rho \pi} \hs{2mm} = \hs{2mm} - \hs{1mm} i \hs{1mm}
{m_{D_s} \hs{1mm} f_{\pi^{'}} \over
m_{\pi^{'}}^2 - m_{D_s}^2
- i \hs{1mm} m_{\pi^{'}} \hs{1mm} \G_{\pi^{'}}(m_{D_s}^2) }
\hs{2mm}
g_{\pi^{'}\rho\pi} \hs{1mm} g_{\rho\pi\pi}
\hs{2mm} = \hs{2mm}
|A_{\rho\pi}| \hs{2mm} e^{i \hs{1mm} \phi_{\rho\pi}}
\label{eq:55}
\eeq
\beq
H_{\rho \pi}(E_1, E_2) \hs{2mm} = \hs{2mm}
{s_2 - s_3 \over m_{\rho}^2 - s_1 - i \hs{1mm} \sqrt{s_1} \hs{1mm} \G_{\rho}}
\hs{2mm} + \hs{2mm}
{s_1 - s_3 \over m_{\rho}^2 - s_2 - i \hs{1mm} \sqrt{s_2} \hs{1mm} \G_{\rho}}
\label{eq:56}
\eeq
For the $f_0\pi$ component which is of the spectator type we obtain, from
Eqs.(\ref{eq:36}) and (\ref{eq:38}),
the structure function $F_{f_0\pi}(E_1, E_2)$ rewritten in terms of the form
factor
$F_0^{D_s f_0}(m_{\pi}^2)$
\beq
F_{f_0\pi}(E_1, E_2) \hs{2mm} = \hs{2mm}
A_{f_0\pi} \hs{2mm} H_{f_0\pi}(E_1, E_2)
\label{eq:57}
\eeq
with
\beq
A_{f_0\pi} \hs{2mm} = \hs{2mm}
\left( {m_{D_s} \over m_{f_0}} \right) \hs{2mm}
\left( {f_{\pi} \over f_{D_s} } \right) \hs{2mm}
\left( 1 - {m_{f_0}^2 \over m_{D_s}^2} \right) \hs{2mm}
F_0^{D_s f_0}(m_{\pi}^2) \hs{1mm}
g_{f_0\pi^{+}\pi^{-}}
\hs{2mm} = \hs{2mm}
\left| A_{f_0\pi} \right| \hs{2mm} e^{i \hs{1mm} \phi_{f_0\pi}}
\label{eq:58}
\eeq
\beq
H_{f_0\pi}(E_1, E_2) \hs{2mm} = \hs{2mm}
{m_{f_0}^2 \over m_{f_0}^2 - s_1 - i \hs{1mm} \sqrt{s_1} \hs{1mm} \G_{f_0}}
\hs{2mm} + \hs{2mm}
{m_{f_0}^2 \over m_{f_0}^2 - s_2 - i \hs{1mm} \sqrt{s_2} \hs{1mm} \G_{f_0}}
\label{eq:59}
\eeq
In the spectator model the form factor is real,
such that $\phi_{f_0\pi} = 0 $ or $\pi$.

The total structure function $F(E_1, E_2)$ has three components
\beq
F(E_1, E_2) \hs{2mm} = \hs{2mm}
F_{\rho\pi}(E_1, E_2) + F_{f_0\pi}(E_1, E_2) + F_{NR}
\label{eq:60}
\eeq
and the computation of the total rate is made using the formula
\beq
\cB(D_s^{+} \ra \pi^{+}\pi^{-}\pi^{+}) \hs{2mm} = \hs{2mm}
1.0399 \hs{1mm} \cdot \hs{1mm} 10^{-2} \hs{2mm} I
\label{eq:61}
\eeq
where $I$ is the phase space integral defined in Eqs.(\ref{eq:51}).

Let us first consider the phase space integrals involving
the functions $H(E_1, E_2)$.
We have previously obtained
\beq
{1 \over m_{D_s}^2} \hs{1mm} \int \int
dE_1 \hs{1mm} dE_2 \hs{2mm} = \hs{2mm} 0.1053
\label{eq:62}
\eeq
\beq
{1 \over m_{D_s}^2} \hs{1mm} \int \int \hs{1mm}
| H_{f_0\pi}(E_1, E_2)|^2 \hs{1mm}
dE_1 \hs{1mm} dE_2
\hs{2mm} = \hs{2mm} K \hs{2mm} = \hs{2mm} 5.7217
\label{eq:63}
\eeq
\beq
{1 \over m_{D_s}^2} \hs{1mm} \int \int \hs{1mm}
| H_{\rho\pi}(E_1, E_2)|^2 \hs{1mm}
dE_1 \hs{1mm} dE_2
\hs{2mm} = \hs{2mm} J \hs{2mm} = \hs{2mm} 8.0997
\label{eq:64}
\eeq

For the interferences between the three components we need three more integrals
\beqa
H_{f_0\pi} \hs{2mm} &=& \hs{2mm}
{1 \over m_{D_s}^2} \hs{1mm} \int \int \hs{1mm}
H_{f_0 \pi}(E_1, E_2) \hs{1mm}
dE_1 \hs{1mm} dE_2
\hs{2mm} = \hs{2mm}
|H_{f_0 \pi}| \hs{2mm} e^{i \hs{1mm} \phi_{Hf_0\pi}}
\label{eq:65} \\
H_{ \rho^0\pi} \hs{2mm} &=& \hs{2mm}
{1 \over m_{D_s}^2} \hs{1mm} \int \int \hs{1mm}
H_{ \rho \pi}(E_1, E_2) \hs{1mm}
dE_1 \hs{1mm} dE_2
\hs{2mm} = \hs{2mm}
\left| H_{ \rho \pi} \right| \hs{2mm} e^{i \hs{1mm} \phi_{H\rho\pi}}
\label{eq:66} \\
H_{ \rho f} \hs{2mm} &=& \hs{2mm}
{1 \over m_{D_s}^2} \hs{1mm} \int \int \hs{1mm}
H_{ \rho\pi}^{*}(E_1, E_2)
H_{f_0 \pi}(E_1, E_2) \hs{1mm}
dE_1 \hs{1mm} dE_2
\hs{2mm} = \hs{2mm}
|H_{\rho f}| \hs{2mm} e^{i \hs{1mm} \phi_{H\rho f}}
\label{eq:67}
\eeqa
The numerical results are
\beqa
| H_{f_0 \pi} | \hs{2mm} &=& \hs{2mm} 0.2677 \hs{30mm}
\phi_{Hf_0 \pi} \hs{2mm} = \hs{2mm} 89.6^0
\label{eq:68} \\
| H_{ \rho \pi} | \hs{2mm} &=& \hs{2mm} 5.58 \cdot 10^{-5} \hs{22mm}
\phi_{H \rho \pi} \hs{2mm} = \hs{2mm} -84.5^0
\label{eq:69} \\
| H_{\rho f} | \hs{2mm} &=& \hs{2mm} 0.3868 \hs{30mm}
\phi_{H \rho f} \hs{2mm} = \hs{2mm} 241.5^0
\label{eq:70}
\eeqa

The numerical values of $|F_{NR}|$ and $|A_{f_0 \pi}|$
are taken from the measured rates respectively in Eqs.(\ref{eq:8})
and (\ref{eq:9}).
In the $\rho\pi$ case we write $A_{\rho\pi} = \lambda \hs{1mm} A_{\rho
\pi}^{max}$
with $0 \leq \lambda < 1$
where $\lambda = 1 $ corresponds to the $90 \%$ C. L. experimental limit
in Eq.(\ref{eq:10}) .

There is \`a priori no reason for the non resonant amplitude $F_{NR}$
to be real and we allow an arbitrary phase $\phi_{NR}$.
The two other phases $\phi_{f_0\pi}$ and $\phi_{\rho\pi}$ are
determined by our model as
\beqa
\phi_{\rho \pi} &=& (70 \pm 10)^0 \hs{17mm}
for \hs{5mm} \G_{\pi^{'}} \hs{2mm} = \hs{2mm} (400 \pm 200) MeV
\label{eq:71} \\
\phi_{f_0 \pi} &=& 0^0 \hs{31mm}
for \hs{10mm} F_0^{D_s f_0}(m_{\pi}^2) \hs{2mm} > \hs{2mm} 0.
\label{eq:72}
\eeqa
The formula for $I(\pi^{+}\pi^{-}\pi^{+})_{TOT}$ is then written as
\beqa
I(\pi^{+}\pi^{-}\pi^{+})_{TOT} \hs{2mm} &=& \hs{2mm}
I(\pi^{+}\pi^{-}\pi^{+})_{NR} \hs{2mm} + \hs{2mm}
I(\pi^{+}\pi^{-}\pi^{+})_{f_0\pi} \hs{2mm} + \hs{2mm}
\lambda^2 \hs{2mm} I(\pi^{+}\pi^{-}\pi^{+})_{\rho \pi}^{max} \hs{2mm}
\no \\
&+& \hs{2mm} 2 \hs{1mm} |F_{NR}| \hs{1mm} |A_{f_0\pi}| \hs{1mm}
|H_{f_0 \pi}| \hs{1mm}
Cos[ \phi_{f_0\pi} + \phi_{Hf_0\pi} - \phi_{NR} ] \no \\
&+& \hs{2mm} 2 \hs{1mm} \lambda \hs{1mm} |A_{\rho\pi}|^{max} \hs{1mm}
|A_{f_0\pi}| \hs{1mm} |H_{\rho f}| \hs{1mm}
Cos[ \phi_{f_0\pi} - \phi_{\rho\pi} + \phi_{H\rho\pi} ] \label{eq:73} \\
&+&  \hs{2mm} 2 \hs{1mm} \lambda \hs{1mm} |F_{NR}| \hs{1mm}
|A_{\rho\pi}|^{max} \hs{1mm} |H_{\rho \pi}| \hs{1mm}
Cos[ \phi_{\rho\pi} + \phi_{H\rho\pi} - \phi_{NR} ] \no
\eeqa
The errors for $I(\pi^{+}\pi^{-}\pi^{+})_{TOT}$ are computed in quadrature
from the experimental errors on $I(\pi^{+}\pi^{-}\pi^{+})_{NR}$ and
$I(\pi^{+}\pi^{-}\pi^{+})_{f_0\pi}$
and for the errors of the amplitudes $|F_{NR}|$ and $|A_{f_0\pi}|$,
we use the approximate relation $\sam x \simeq {1 \over 2 x} \sam x^2$.

The result of our calculation of $I(\pi^{+}\pi^{-}\pi^{+})_{TOT}$
directly gives the total branching ratio by the relation (\ref{eq:61})
and is represented in Figs.3
as a function of the phase parameter $\phi_{NR}$
for the two extreme cases $\lambda = 0$
and $\lambda = 1$.
The one standard deviation domains can be compared with
the experimental result of  Eq.(\ref{eq:7}).

It is clear from  Fig.3-a corresponding to a two component model
$(\ld = 0)$, and
Fig.3-b where a maximal $\rho^0\pi^{+}$ contribution has been added
$(\ld = 1)$, that
there is no difficulty in fitting the experimental result (\ref{eq:7})
within one standard deviation, for any value of the phase parameter
$\phi_{NR}$.

\vs{7mm}
\hs{3mm} \large{ } {\bf VI. \hs{3mm}  Energy distributions of the pion
in $D_s^{+} \ra \pi^{+} \pi^{+} \pi^{-} $ decay }
\normalsize
\vs{5mm}

\large
1). \normalsize
The two $\pi^{+}$ meson energies being $E_1$ and $E_2$
with our convention of section II, the function $F(E_1, E_2)$
introduced in section V is symmetrical
in the exchange of $E_1$ and $E_2$.
Such a property extends to the Dalitz plot which is a double
differential distribution proportional to $|F(E_1, E_2)|^2$.
However it is probably premature to discuss the detailed properties of the
Dalitz plot and it is more realistic to consider single differential
quantities like the $\pi$ meson energy distributions.

We must keep in mind that a distribution in the pion energy $E_j$ is just
the mirror of a distribution in the invariant two pion mass squared $s_j$
because of the relations (\ref{eq:1}),
$s_j = m_{D_s}^2 + m_{\pi}^2 - 2 m_{D_s} E_j$.

\large
2). \normalsize
For the final state $\pi^{+}\pi^{+}\pi^{-}$, we define the $\pi^{+}$
and $\pi^{-}$  energy distributions in the following way
\beq
{d\G(D_s^{+} \ra \pi^{+}\pi^{+}\pi^{-}) \over dE_1}
\hs{2mm} = \hs{2mm}
\tilde{\cN} \hs{1mm} G_{+}(E_1) \hs{7mm}, \hs{7mm}
{d\G(D_s^{+} \ra \pi^{+}\pi^{+}\pi^{-}) \over dE_3}
\hs{2mm} = \hs{2mm}
\tilde{\cN} \hs{1mm} G_{-}(E_3)
\label{eq:76}
\eeq
where the distributions $G_{+}(E_1)$ and $G_{-}(E_3)$ are defined by
\beqa
G_{+}(E_1) \hs{2mm} &=& \hs{2mm}
{1 \over m_{D_s}} \hs{1mm} \int_{E_{-}(E_1)}^{E_{+}(E_1)} \hs{1mm}
|F(E_1, E_2)|^2 \hs{2mm} dE_2
\label{eq:77} \\
G_{-}(E_3) \hs{2mm} &=& \hs{2mm}
{1 \over m_{D_s}} \hs{1mm} \int_{E_{-}(E_3)}^{E_{+}(E_3)} \hs{1mm}
|F(m_{D_s} - E_2 - E_3, E_2)|^2 \hs{2mm} dE_2
\label{eq:78}
\eeqa
and the limits of integration $E_{\pm}(E)$ have been given in
Eq.(\ref{eq:5}) of section II.

The normalization constants $\tilde{\cN}$ and $\cN$
are simply related by
\beq
\cN \hs{2mm} = \hs{2mm}
{ \tau_{D_s} \over \hbar } \hs{2mm} m_{D_s} \hs{2mm} \tilde{\cN}
\label{eq:79}
\eeq
and the numerical value of $\tilde{\cN}$
is $\tilde{\cN} = 7.446 \cdot 10^{-15}$.

The $\pi^{+}$ energy distribution $G_{+}(E_1)$ and the $\pi^{-}$
energy distribution $G_{-}(E_3)$ for all $\pi^{+}\pi^{+}\pi^{-}$
events have been computed performing a single variable integration
with the relevant functions defined in section V,
$|H_{f_0\pi}(E_1,E_2)|^2$, $|H_{\rho\pi}(E_1,E_2)|^2$,
$H_{f_0\pi}(E_1,E_2)$, $H_{\rho\pi}(E_1,E_2)$, and
$H^{*}_{\rho\pi}(E_1,E_2)H_{f_0\pi}(E_1,E_2)$.
We observe that the last three integrals are complex with
an energy dependent phase and, as a consequence, the nature of constructive
or destructive of the various interferences is an energy dependent concept.
The terms $F_{NR}$, $A_{f_0\pi}$, and $A_{\rho\pi}$ are
treated as in section V with their moduli extracted
from the measured branching ratios.
The two extreme cases $\lambda = 0$ and $\lambda = 1$ have been considered.

It is clear that the shape of the $\pi$ meson energy distributions
$G_{+}(E_1)$ and $G_{-}(E_3)$ depends on the constant phase of
the non resonant amplitude $F_{NR}$ in a very sensitive way.
We notice in \cite{R1} that
we have a Dalitz plot in the $(s_1, s_2)$ plane corresponding
to the $(68.1 \pm 12.4)$ events observed in
$D_s^{+} \ra \pi^{+}\pi^{-}\pi^{+}$ decays.
A projection of this Dalitz plot on the $s_1$ axis shows
clearly a peak associated to the $f_0\pi^{+}$ quasi two body
 state and an important background.
Of course, such an histogram in \cite{R1} is nothing but
our $\pi^{+}$ meson energy distribution $G_{+}(E_1)$.

The $G_{+}(E_1)$ distribution has been studied for values of
$\phi_{NR}$ between $0^0$ and $360^0$.
Comparing our theoretical curves with
the  shape of the experimental histogram \cite{R1},
we obtain an acceptable agreement
if the angle $\phi_{NR}$ is in the range
\beq
70^0 \hs{2mm} < \hs{2mm} \phi_{NR} \hs{2mm} < \hs{2mm} 140^0
\label{eq:80}
\eeq

Let us now come back to Eq.(\ref{eq:48}) where the function
$F(E_1, E_2)$ is defined.
Beside the phase of the CKM factors, we have the phase $\pi/2$ due to
the factor $i$ entering in the definition of the matrix element
$< 0 \hs{1mm} | \cA_{\mu} |D_s^{+} >$
and the phase of the function $F(E_1, E_2)$.
In the non resonant component, the last phase is $\phi_{NR}$.
We observe that the value $\phi_{NR} = 90^0$ compatible with the range
(\ref{eq:80}) makes the quantity $i \hs{1mm} F_{NR}$ real and negative.
In the absence of more detailed experimental informations
we shall make a simple choice $\phi_{NR} = 90^0$ in what follows.

We have represented in Figs.4 and Figs.5 the pion energy distributions
$G_{+}(E_1)$ and $G_{-}(E_3)$ for the two cases $\lambda = 0$ and
$\lambda = 1$ using $\phi_{NR} = 90^0$.
For completeness we give the theoretical branching ratios corresponding to
$\phi_{NR} = 90^0$ which can be directly extracted from Figs.3.
\beqa
(i) \hs{10mm} \lambda = 0, & & \hs{15mm}
\cB(D_s^{+} \ra \pi^{+} \pi^{-} \pi^{+})_{TH} \hs{2mm} = \hs{2mm}
2.022 \pm 0.528 \hs{2mm} \%,
\label{eq:81} \\
(ii) \hs{10mm} \lambda = 1, & & \hs{15mm}
\cB(D_s^{+} \ra \pi^{+} \pi^{-} \pi^{+})_{TH} \hs{2mm} = \hs{2mm}
2.259 \pm 0.524 \hs{2mm} \%.
\label{eq:82}
\eeqa
Both values Eqs.(\ref{eq:81}) and (\ref{eq:82}) agree with the experimental one
of Eq.(\ref{eq:7}), $(1.54 \pm 0.42) \hs{2mm} \%$.

\vs{7mm}
\hs{3mm} \large{ }
{\bf VII. \hs{3mm} Predictions for $D_s^{+} \ra \pi^{0} \pi^{0}
\pi^{+} $ decay mode.}
\normalsize
\vs{5mm}

We have no experimental data on the decay mode
$D_s^{+} \ra \pi^0 \pi^0 \pi^{+}$.
However we shall use the same three component model for this decay mode
and let us start our discussion with isospin considerations
concerning these three components.

\large
1). \normalsize
Consider first the case of a non resonant $\pi^0 \pi^0 \pi^{+}$ state.
The constant $F_{NR}$ has no reason to be the same for the two modes
$\pi^{+} \pi^{-} \pi^{+}$ and $\pi^0 \pi^0 \pi^{+}$.
By assumption, with a constant function $F_{NR}(E_1, E_2)$,
we have a full symmetry in space between the three pions.
{}From the Bose-Einstein symmetry,
the isospin configuration has also to be totally symmetric.
Consider now a third rank fully symmetric tensor in a three dimensional space.
It has 10 independent components.
With respect to the isospin SO(3) orthogonal group,
such a tensor is reducible
into an isospin $I = 3$ part with 7 components
and an isospin $I = 1$ part with 3 components.
In our specific $W$ annihilation model for the non resonant three pions,
only the later part contributes,
the $u\ol{d}$ weak current being an isovector.
By inspection of the relevant Clebsch-Gordan coefficients, we get
the result
\beq
F_{NR}(\pi^{+}\pi^{+}\pi^{-})
= 3 \hs{2mm} F_{NR}(\pi^{0}\pi^{0}\pi^{+}) \hs{3mm}. \label{eq:7-1}
\eeq
As a consequence, in our model, we obtain
for the non resonant part :
\beq
\frac{ \G(D_s^{+} \ra \pi^0 \pi^0 \pi^{+})_{NR}}
{ \G(D_s^{+} \ra \pi^{+} \pi^{+} \pi^{-})_{NR}}
\hs{2mm} = \hs{2mm} {1 \over 9}
\hs{2mm} = \hs{2mm} 11.1 \hs{2mm} \% \hs{3mm}, \label{eq:83}
\eeq
and the non resonant branching ratio
$\cB(D_s^{+} \ra \pi^0 \pi^0 \pi^{+})_{NR}$ is
expected to occur only at the $10^{-3}$ level.

\large
2). \normalsize
The quasi two body state $\rho^0\pi^{+}$ cannot produce a
$\pi^0\pi^0\pi^{+}$ final state because of the isovector character
of the $\rho$ meson.
However $\rho^{+}\pi^0$ can give such a state when followed by the decay
$\rho^{+} \ra \pi^{+}\pi^{0}$.
The $D_s^{+}$ meson being an isoscalar, the decay amplitudes for
$D_s \ra \rho^0 \pi^{+}$ and $D_s \ra \rho^{+}\pi^0$ are equal.
When mass differences between charged and neutral $\rho$'s and $\pi$'s
are neglected we obtain
\beq
\G(D_s^{+} \ra \rho^0 \pi^{+})
\hs{2mm} = \hs{2mm}
\G(D_s^{+} \ra \rho^{+} \pi^0)
\label{eq;84}
\eeq
With equal decay widths for
$\rho^{0} \ra \pi^{+}\pi^{-}$ and $\rho^{+} \ra \pi^{+}\pi^{0}$
we obtain the ratio of rates
\beq
{ \G(D_s^{+} \ra \rho^{+} \pi^0 \ra \pi^{+} \pi^{0} \pi^{0}) \over
\G(D_s^{+} \ra \rho^{0} \pi^+ \ra \pi^{+} \pi^{-} \pi^{+}) }
\hs{2mm} = \hs{2mm} 1
\label{eq:85}
\eeq
We observe that the finite $\rho$ width corrections
are the same in both final states $\pi^{+} \pi^{-} \pi^{+}$
and $\pi^{0} \pi^{0} \pi^{+}$
because two Breit Wigner amplitudes are present in the variables
$s_1$ and $s_2$ with the choice made in section II.

\large
3). \normalsize
For the third component, i.e., the decay $D_s^{+} \ra f_0 \pi^{+}$
followed by  $f_0 \ra \pi^{+}\pi^{-}$ $(f_0 \ra \pi^0 \pi^0)$
produces a $\pi^{+}\pi^{-}\pi^{+}(\pi^0\pi^0\pi^{+})$ final state.
Because of the isoscalar character of the $f_0(980)$ we have
\beq
\G(f_0 \ra \pi^{+}\pi^{-}) \hs{2mm} = \hs{2mm}
2 \hs{2mm} \G(f_0 \ra \pi^0\pi^0)
\label{eq:86}
\eeq
On the other hand, in the $D_s^{+} \ra f_0\pi^{+} \ra \pi^{+}\pi^{-}\pi^{+}$
case, we have two Breit-Wigner amplitudes in the variables $s_1$ and $s_2$
whereas in the
$D_s^{+} \ra f_{0} \pi^{+} \ra \pi^0 \pi^0 \pi^{+}$ case
there exists only one Breit-Wigner amplitude in the variable $s_3$
with the choice made in section II.
As a consequence the phase space integral $K$ introduced in
Eq.(\ref{eq:40}) has the value $K_c = 5.7217$
for the final state $\pi^+ \pi^- \pi^{+}$ and
$K_N = 2.7469$
for the final state $\pi^0 \pi^0 \pi^{+}$, and
we obtain the ratio of rates
\beq
{ \G(D_s^{+} \ra f_0 \pi^+ \ra \pi^{0} \pi^{0} \pi^{+}) \over
\G(D_s^{+} \ra f_0 \pi^+ \ra \pi^{+} \pi^{-} \pi^{+}) }
\hs{2mm} = \hs{2mm}
{1 \over 2} \hs{2mm} {K_N \over K_c}
\hs{2mm} = \hs{2mm}
24 \%
\label{eq:87}
\eeq
The departure of this ratio from ${1 \over 4}$ is simply
due to the interference between the two Breit-Wigner contributions
in the $D_s^{+} \ra  f_0\pi^{+} \ra \pi^{+}\pi^{-}\pi^{+}$ case.

\large
4). \normalsize
The function $F(E_1,E_2)$ associated to the final state
$\pi^0\pi^0\pi^{+}$ is written in the form
\beq
F(E_1, E_2) \hs{2mm} = \hs{2mm}
{1 \over 3} F_{NR} \hs{2mm} + \hs{2mm}
{1 \over \sqrt{2}} A_{f_0\pi} \hs{1mm} H_{f_0\pi}(E_1, E_2)
\hs{2mm} + \hs{2mm}
\lambda \hs{1mm} A_{\rho\pi}^{max} \hs{1mm} H_{\rho\pi}(E_1, E_2)
\label{eq:88}
\eeq
where in the right hand side of Eq.(\ref{eq:88})
all terms but $H_{f_0\pi}(E_1,E_2)$ are the same as
those defined in section V for the $\pi^{+}\pi^{-}\pi^{+}$
final state.
In Eq.(\ref{eq:88}), $E_1$ and $E_2$ are the $\pi^0$
energies and $E_3$ the $\pi^{+}$ energy.
Whereas $H_{f_0\pi}(E_1, E_2)$ in the $\pi^{+}\pi^{-}\pi^{+}$ case
contains two Breit-Wigner terms with the variables $s_1$ and $s_2$
(Eq.(\ref{eq:59}) ), for the $\pi^0\pi^0\pi^{+}$ final state
we have only one Breit-Wigner term with the variable $s_3$
\beq
H_{f_0\pi}(E_1, E_2) \hs{2mm} = \hs{2mm}
{m_{f_0}^2 \over m_{f_0}^2 - s_3 - i \hs{1mm} \sqrt{s_3} \hs{1mm} \G_{f_0}}
\label{eq:89}
\eeq

The total rate is now computed by integration
the quantity $|F(E_1, E_2)|$ over $E_1$ and $E_2$.
The procedure of calculation is similar to the one explained
in detail in section V
and the result for $I(\pi^0\pi^0\pi^{+})_{TOT}$ is shown on Figs.6
for the two extreme cases $\lambda = 0$ and $\lambda = 1$.
Because of the isospin factor ${1 \over 3}$ in front of $F_{NR}$
the role played by the non resonant component is obviously
less important in the $\pi^0\pi^0\pi^{+}$ case than it was in the
$\pi^+\pi^-\pi^+$ one.
For such a reason the quantity $I(\pi^0\pi^0\pi^+)_{TOT}$
in the two component model, $\lambda = 0$,
is insensitive to the phase parameter $\phi_{NR}$.
Again due to the isospin factors, the possible role of
the quasi two body state $\rho\pi$ might be more important
in the $\pi^0\pi^0\pi^{+}$ case than it was
for the $\pi^+\pi^-\pi^+$ one.
Such a qualitative expectation is clearly seen in Figs.6.
Both Figs.3 and 6 are drawn at the same scale
and the order of magnitude of the branching ratio
$\cB(D_s^{+} \ra \pi^0\pi^0\pi^+)$ is expected to be few
$10^{-3}$ and that might be the reason why the mode
$D_s^{+} \ra \pi^0\pi^0\pi^+$ has not yet been experimentally observed.

With the choice $\phi_{NR} = 90^0$ we find
\beqa
(i) \hs{5mm} \lambda &=& 0, \hs{20mm}
\cB(D_s^+ \ra \pi^0\pi^0\pi^{+})_{TH} \hs{2mm} = \hs{2mm}
0.294 \pm 0.092 \hs{2mm} \%,
\label{eq:90} \\
(ii) \hs{5mm} \lambda &=& 1, \hs{20mm}
\cB(D_s^+ \ra \pi^0\pi^0\pi^{+})_{TH} \hs{2mm} = \hs{2mm}
0.604 \pm 0.101 \hs{2mm} \%.
\label{eq:91}
\eeqa

\large
5). \normalsize
The $\pi^0$ and $\pi^+$ energy distributions $G_0(E_1)$ and $G_{+}(E_3)$
for the final state $\pi^0\pi^0\pi^{+}$
are computed with the formulae (\ref{eq:77}) and (\ref{eq:78})
using now the function $F(E_1, E_2)$ given in Eq.(\ref{eq:88}).
The procedure of calculation is similar to the one explained in section VI
for the $\pi^+\pi^-\pi^+$ final state.

We have represented respectively in Figs.7 and 8
\hs{1mm} the $\pi^0$ and $\pi^+$
energy distributions $G_0(E_1)$ and $G_{+}(E_3)$ for the two extreme situations
$\lambda = 0$ and $\lambda = 1$, using as previously the phase
$\phi_{NR} = 90^0$ for the non resonant amplitude.
 Figs.8 show clearly the peak
due to the quasi two body $f_0\pi^{+}$ state
and a possible quasi two body $\rho^+\pi^0$ contribution
is clearly seen in Fig.7-b.

Let us end by considering the $\pi$ meson energy distribution
corresponding to an uniformly populated phase space of a constant amplitude.
The energy distributions for each pion are obviously identical
and in the case
$D_s^+ \ra (\pi^+\pi^-\pi^+)_{NR}$ we obtain, using Eq.(\ref{eq:5}),
the analytic form
\beq
G_{NR}(E) \hs{2mm} = \hs{2mm}
|F_{NR}|^2 \hs{2mm}
\left\{ {(E^2 - m_{\pi}^2)(m_{D_s}^2 - 3 m_{\pi}^2 - 2 m_{D_s}E) \over
m_{D_s}^2 + m_{\pi}^2 - 2 m_{D_s}E }
\right\}^{1/2}
\label{eq:93}
\eeq
Using the experimental value (\ref{eq:53}) of $|F_{NR}|^2$,
we obtain the distribution represented in Fig.9.

\vs{7mm}
\hspace{3mm} \large{} {\bf VIII. \hs{3mm} Summary and Concluding Remarks}
\vspace{5mm}
\normalsize

We have studied, in this paper, the decay of the
$D_s^{+}$ meson into three $\pi$ mesons by analyzing
the experimental data available for the decay mode
$D_s^{+} \ra \pi^+\pi^-\pi^+$.
The amplitude can be considered
as the superposition of two mechanisms :
the spectator decay of the $c$ quark contained in the $D_s^{+}$ meson
and the $W^{+}$ annililation into three pions.
Both mechanisms are needed to explain the data
and the latter  appears to be considerably important.

The non observation of the $\rho\pi$ quasi two body state
allows us to determine an upper limit of about $10 \hs{2mm} MeV$
for the leptonic decay constant
$f_{\pi^{'}}$ of the $\pi^{'} \equiv \pi(1300)$ meson.
Such an upper bound  of $f_{\pi^{'}}$ is compatible with
theoretical estimates using non-relativistic chiral quark model \cite{R9}
and $QCD$ sum rule technique \cite{R10}.
We remark that the commonly used values of the $f_{\pi^{'}}$ decay constant
\cite{R6} are much larger than our upper limit.

The large $f_0 \pi^{+}$ branching ratio
compared to the $\rho^0 \pi^+$ one allows us to exclude
the dominant $(u\ol{u} + d\ol{d})/\sqrt{2}$
quark structure for the $f_0(980)$ meson.
 Assuming the $f_0(980)$ to be a $s\ol{s}$ state,
the spectator mechanism will govern the decay
$D_s^{+} \ra f_0(980) \pi^+$ and from experimental data it is
possible to estimate the hadronic form factor
$F_0^{D_sf_0}(m_{\pi}^2) $ to be
$0.36 \hs{1mm}^{+ \hs{1mm} 0.06}_{- \hs{1mm} 0.08}$.

Furthermore,  a large branching ratio of the $D_s^{+}$ meson
into a non resonant three pion state
has been experimentally measured.
This decay can only proceed through the WA
and the corresponding decay amplitude is assumed to be independent
of the $\pi$ meson energies.
We then introduce a phenomenological complex constant $F_{NR}$,
its modulus is extracted from the non resonant branching ratio
$\cB(D_s^{+} \ra \pi^+\pi^-\pi^+)_{NR}$.

For the full decay amplitude, we propose a three component model involving
the non resonant background, the $f_0\pi^{+}$ quasi two body state, and
a possible $\rho\pi$ quasi two body state with a parameter $\lambda$
such that for $\lambda = 1 $, the branching ratio
$\cB(D_s^{+} \ra \rho^0\pi^{+})$ takes the value of the $90 \%$
confidence level upper limit.
We first compute the total rate for $D_s^{+} \ra \pi^+\pi^-\pi^+$
as a function of the phase $\phi_{NR}$ of the constant amplitude $F_{NR}$.
Agreement with experiment is obtained,
within one standard deviation, in both cases $\ld = 0$ and $\ld = 1$,
for all values of $\phi_{NR} $ between $0^0$ and $360^0$.

For the decay mode $D_s^+ \ra \pi^+\pi^-\pi^+$, we compute
the $\pi^+$  and $\pi^-$ energy distributions which
depend on the phase parameter  $\phi_{NR}$.
A qualitative agreement between our $\pi^+$ distribution
with the experimental histogram \cite{R1}
in the invariant $(\pi^+\pi^-)$ mass
is obtained when $\phi_{NR}$ is restricted in the range given by
Eq.({\ref{eq:80})
including $\phi_{NR} = 90^0$.
Fixing now $\phi_{NR} = 90^0$,  we present our predictions
for the $\pi^+$ and $\pi^-$ energy distributions including one standard
deviation errors deduced for those of the experimental rates.

The same model is used to make preditions for the decay mode
$D_s^+ \ra \pi^0\pi^0\pi^+$ which is not yet experimentally observed.
The total rate for $D_s^+ \ra \pi^0\pi^0\pi^+$
is found to be essentially one order of magnitude smaller
than the $D_s^+ \ra \pi^+\pi^-\pi^+$ rate
and this might be the reason why the decay
$D_s^+ \ra \pi^0\pi^0\pi^+$ has not yet been detected.
Also the $\pi^0$ and $\pi^+$ energy distributions are predicted by our model.

If the quasi two body states $f_0\pi$ and $\rho\pi$ can be
theoretically understood, the non resonant part has been treated, in this
paper,
only from a purely phenomenological view point,
 using a constant complex parameter $F_{NR}$.
It would be interesting to explain theoretically the experimental values found
for $|F_{NR}|$ and $\phi_{NR}$.
To our knowledge no model has been proposed for such case.

The detailed understanding of the WA
is very important in its own right, and for that purpose, the
$D_s^{+}$, like the $B_c^{+}$, are particularly auspicious for WA
to manifest in its full strength due to both color and CKM favored factors.
In all other cases, i.e., $D^{+}, D^0, B^{+}, B^0,$ and $B_s^0$,
the WA is either
masked by the color and CKM suppressed factors,
or contaminated by the spectator decay mechanism at the quark level,
such that unambiguous WA effect can be hardly isolated.
Therefore experimental and theoretical investigations of the $D_s^+$
( and the $B_c^+$ later) decays into pions,
in both inclusive and exclusive modes, are of great interest.
They are the ideal laboratories for studying the weak annihilation mechanism.

\vspace{10mm}
\hspace{1cm} \Large{} {\bf Acknowledgements}    \vspace{0.5cm}
\normalsize

\vs{2mm}
Y. Y. K would like to thank the Commissariat \`a l'Energie Atomique of France
for financial support
and especially G. Cohen-Tannoudji and J. Ha\"{\i}ssinski
for their encouragements.

%
%
%
%

\vs{10mm}
\newpage
%

\newpage
\section*{Figure captions}
\normalsize
\vspace{0.5cm}

\ben

\item
{\bf Figure 1} : \hs{3mm}
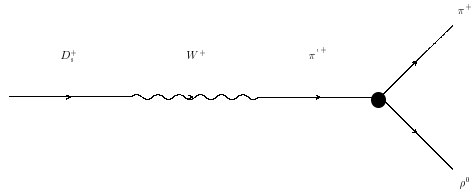
$W$ annihilation mechanism for the $D_s^{+} \ra \rho^0 + \pi^+$ decay mode.

\item
{\bf Figure 2} : \hs{3mm}
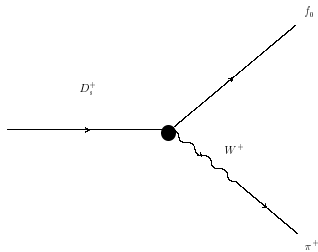
Spectator mechanism for the $D_s^{+} \ra f_0 + \pi^+$ decay mode.

\item
{\bf Figure 3} : \hs{3mm}
Total branching ratio for the decay
$D_s^{+} \ra \pi^+ \pi^- \pi^+$
as a function of the phase parameter $\phi_{NR}$,
(a) in the two component model : $\lambda = 0 $,
(b) in the three component model with the maximal $\rho\pi^+$
contribution : $\lambda = 1$.
One standard deviation errors are indicated and the horizontal bound is
the experimental result including one standard deviation errors.

\item
{\bf Figure 4} :
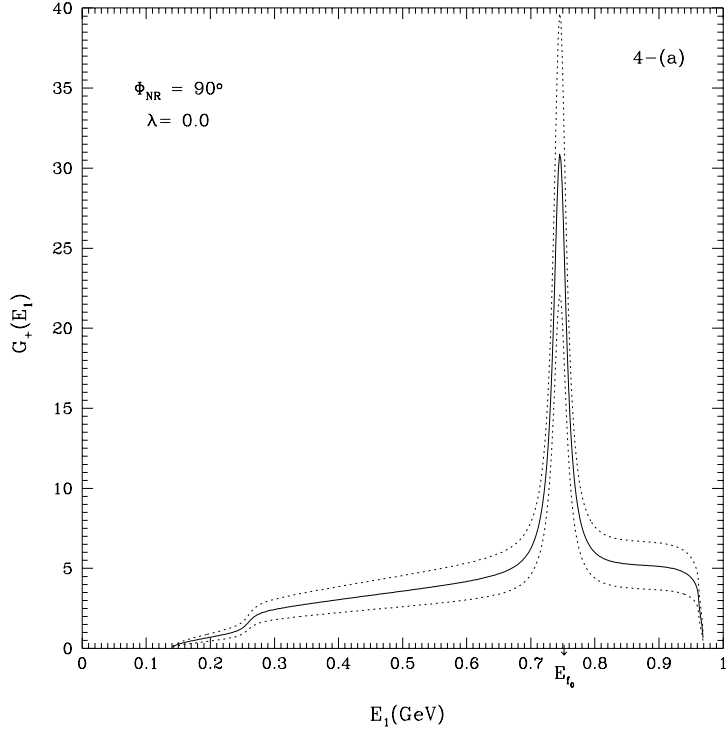
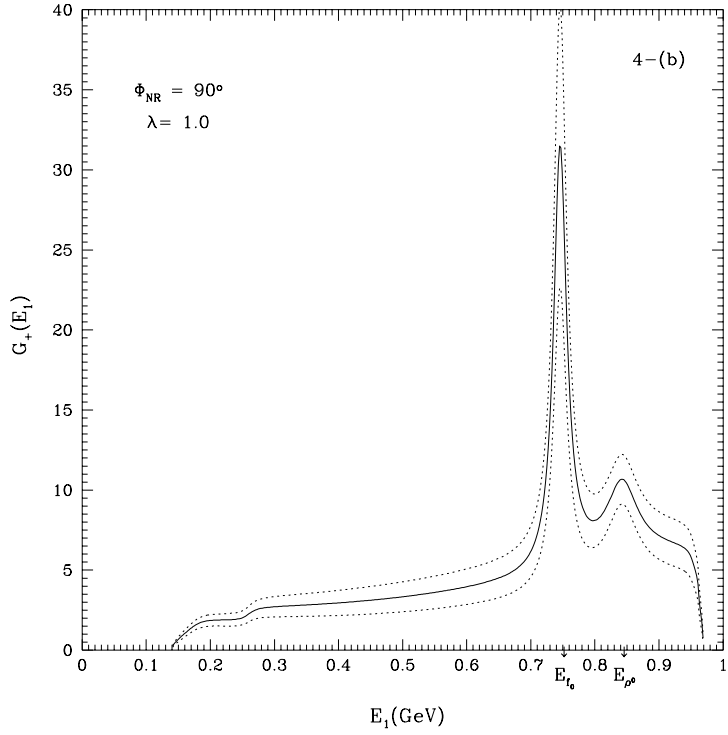
The $\pi^+$ meson energy distribution function $G_{+}(E_1)$ for the final state
$\pi^+\pi^-\pi^+$,
(a) for $\lambda = 0 $ in the two component model,
(b) for $\lambda = 1 $ in the three component model with the maximal
$\rho\pi^+$ contribution.
One standard deviation errors are indicated.
Here the quantity $E_{f_0} = 0.7453 \hs{2mm} GeV$ is associated
to the $f_{0}$ resonance and
$E_{\rho} = 0.8386 \hs{2mm} GeV$ is associated
to the $\rho^0$ resonance.

\item
{\bf Figure 5} :
The $\pi^-$ meson energy distribution function $G_{-}(E_3)$ for the final state
$\pi^+\pi^-\pi^+$,
(a) for $\lambda = 0 $ in the two component model,
(b) for $\lambda = 1 $ in the three component model with the maximal
$\rho\pi^+$ contribution.
One standard deviation errors are indicated.

\item
{\bf Figure 6} :
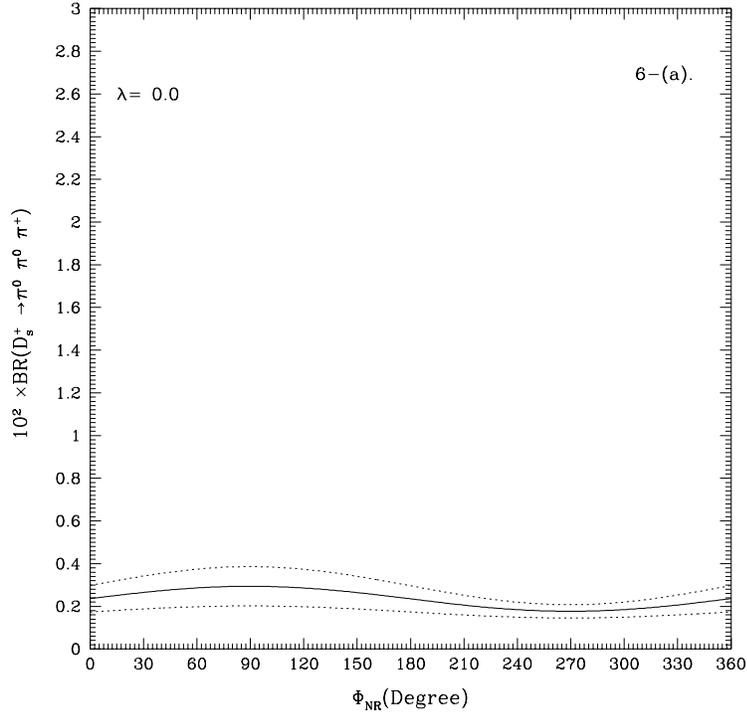
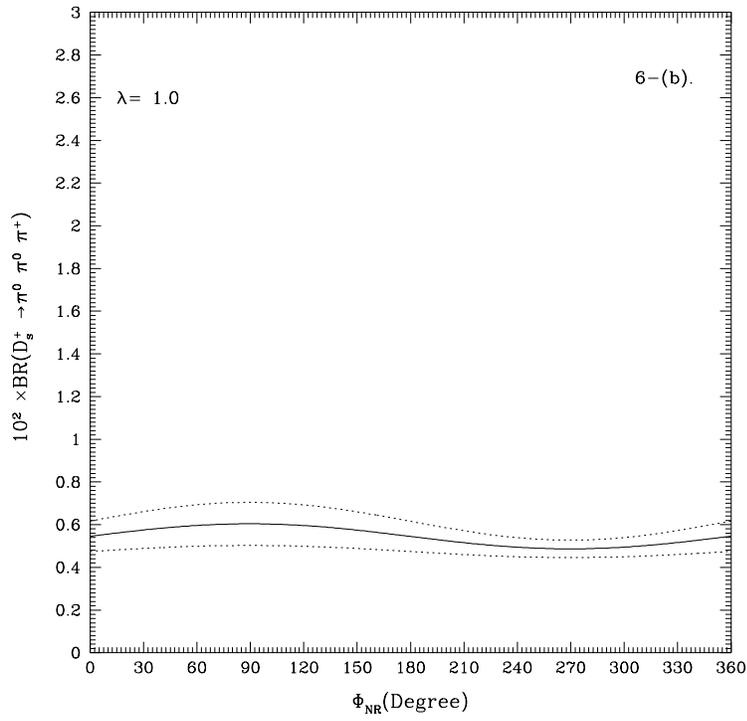
Total branching ratio for the decay
$D_s^{+} \ra \pi^0 \pi^0 \pi^+$
as a function of the phase parameter $\phi_{NR}$,
(a) in the two component model : $\lambda = 0 $,
(b) in the three component model with the maximal $\rho\pi^+$
contribution : $\lambda = 1$.

\item
{\bf Figure 7} :
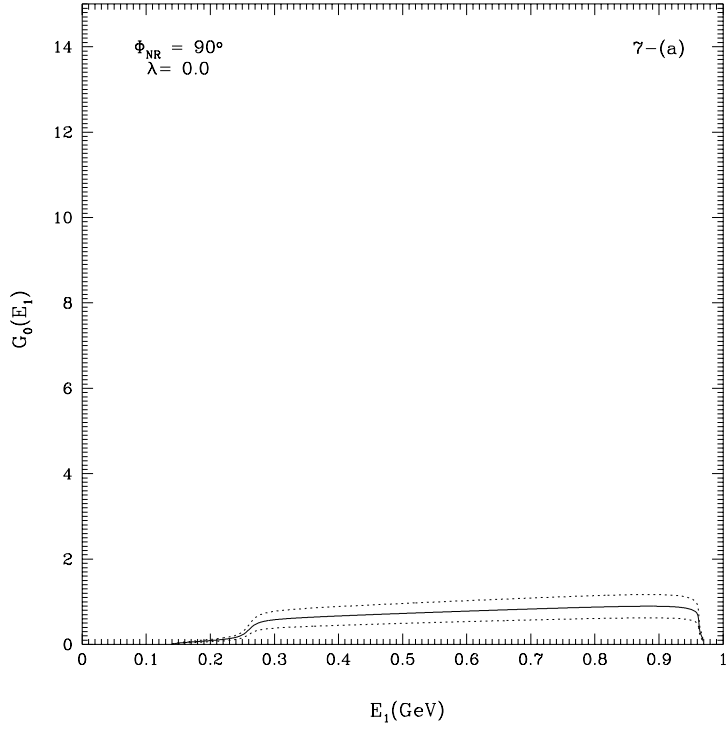
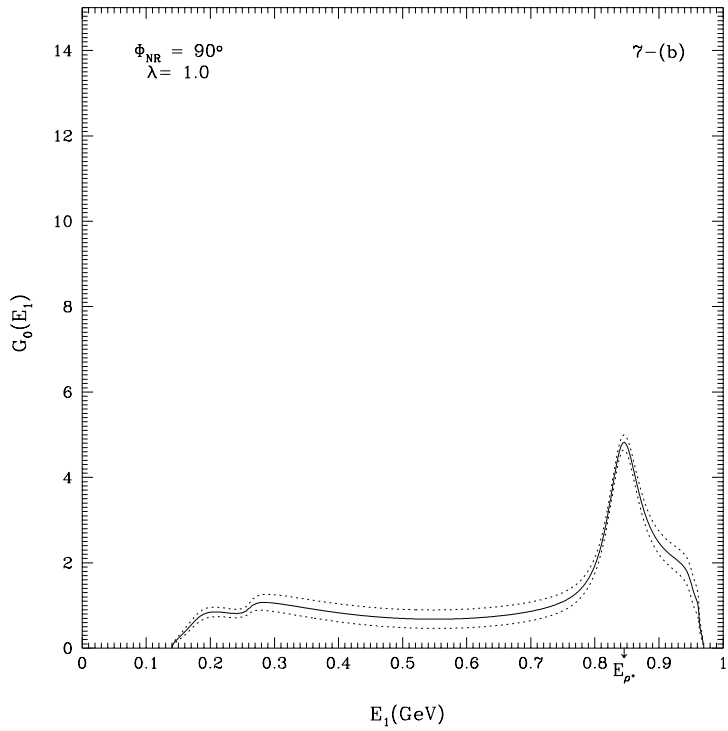
The $\pi^0$ meson energy distribution function $G_{0}(E_1)$ for the final state
$\pi^0\pi^0\pi^+$,
(a) for $\lambda = 0 $ in the two component model,
(b) for $\lambda = 1 $ in the three component model with the maximal
$\rho^+\pi^0$ contribution.
One standard deviation errors are indicated.
Here the quantity
$E_{\rho} = 0.8386 \hs{2mm} GeV$ is associated
to the $\rho$ resonance.

\item
{\bf Figure 8} :
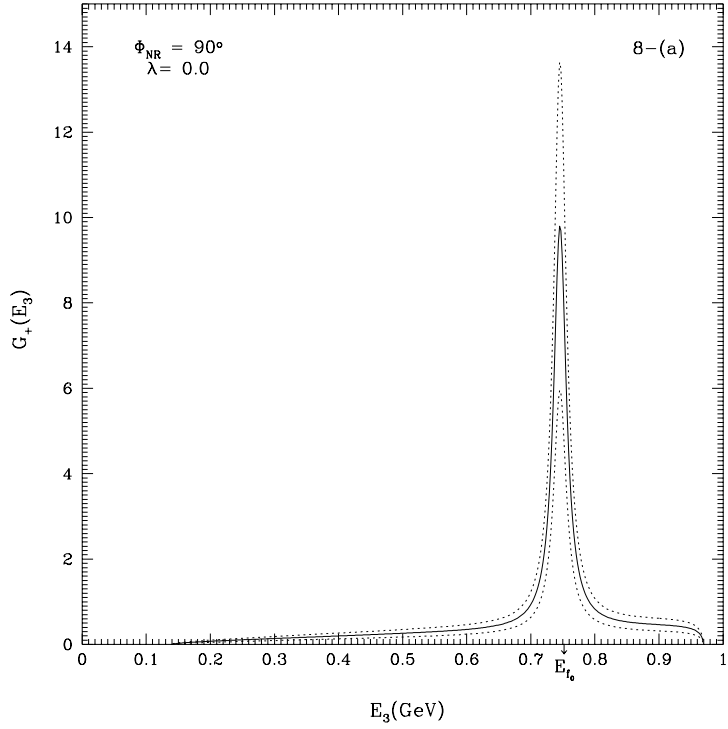
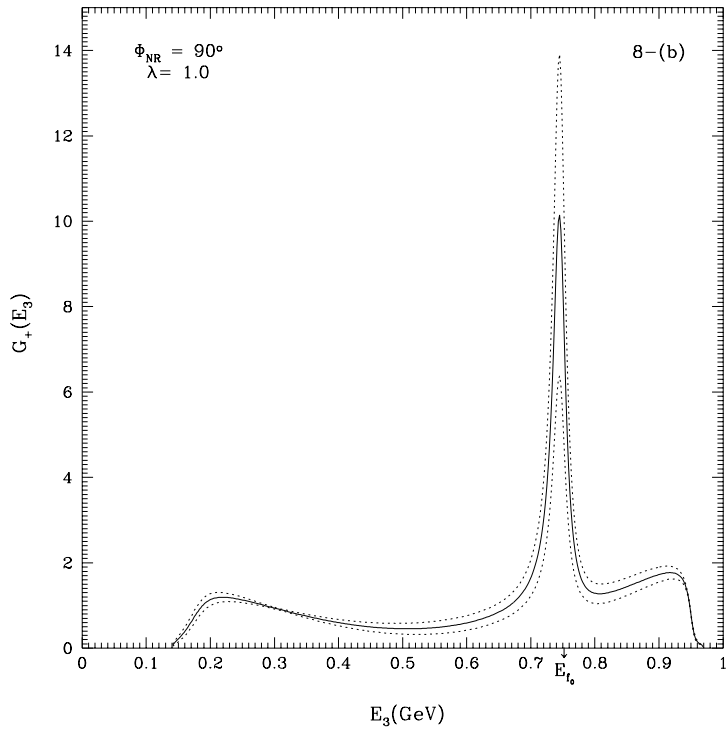
The $\pi^+$ meson energy distribution function $G_{+}(E_3)$ for the final state
$\pi^0\pi^0\pi^+$,
(a) for $\lambda = 0 $ in the two component model,
(b) for $\lambda = 1 $ in the three component model with the maximal
$\rho^+\pi^0$ contribution.
One standard deviation errors are indicated.
Here the quantity $E_{f_0} = 0.7453 \hs{2mm} GeV$ is associated
to the $f_{0}$ resonance.

\item
{\bf Figure 9} : \hs{3mm}
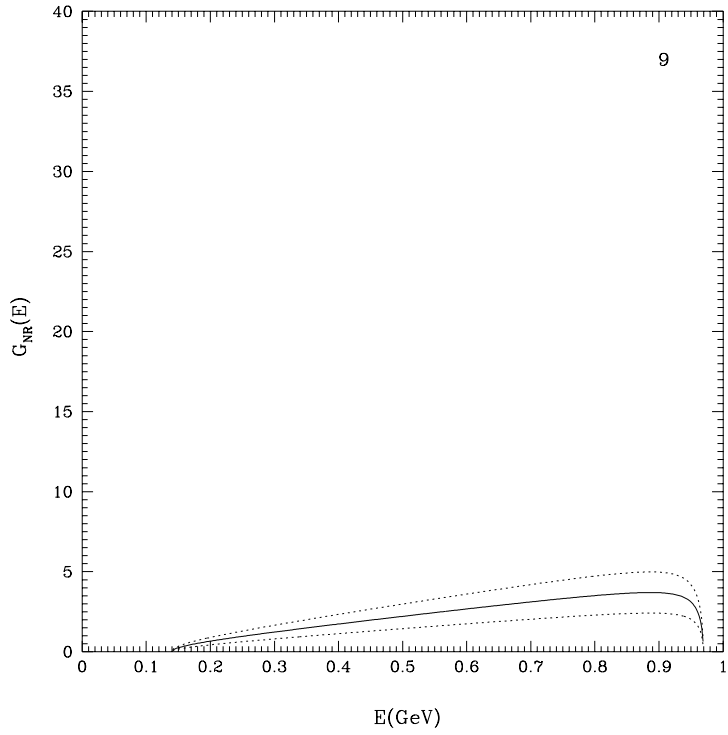
The $\pi$ meson energy distribution $G_{NR}(E)$
for the non resonant part of the decay
$D_s^{+} \ra \pi^+ \pi^- \pi^+$.
One standard deviation errors are indicated.

\een

\end{document}